\documentclass[fleqn,usenatbib]{mnras}
\usepackage[T1]{fontenc}
\usepackage{comment}
\usepackage{graphicx}	
\usepackage{amsmath}
\usepackage{amssymb}
\usepackage{subfigure}
\usepackage{color}

\usepackage{bm}
\DeclareMathAlphabet{\mathsfit}{T1}{\sfdefault}{m}{sl}
\SetMathAlphabet{\mathsfit}{bold}{T1}{\sfdefault}{bx}{sl}
\newcommand\ten[1]   {\bm{\mathsfit{{#1}}}}

\title[]{Ergomagnetosphere, Ejection Disc, Magnetopause in M87.\\ I Global Flow of Mass, Angular Momentum, Energy and Current}

\author[Blandford \& Globus]{Roger Blandford$^{1}$ and No\'emie Globus$^{2,3}$\\
$^{1}$Kavli Institute for Particle Astrophysics and Cosmology (KIPAC), Stanford University, Stanford, CA 94305, USA \\$^2$Department of Astronomy and Astrophysics, University of California, Santa Cruz, CA 95064, USA \\$^3$Center for Computational Astrophysics, Flatiron Institute, Simons Foundation, New-York, NY10003, USA} 
\date{Accepted XXX. Received YYY; in original form ZZZ}
\pubyear{2021}
\begin{document}
\label{firstpage}
\pagerange{\pageref{firstpage}--\pageref{lastpage}}
\maketitle 
\begin{abstract}
We interpret the 1.3mm VLBI observations made by the Event Horizon Telescope of the black hole in M87.  It is proposed that, instead of being a torus of accreting gas, the observed annular ring is a rotating, magnetically-dominated ergomagnetosphere that can transmit electromagnetic angular momentum and energy outward to the disc through a combination of large scale magnetic torque and small scale instabilities. It is further proposed that energy can be extracted by magnetic flux threading the ergosphere through the efficient emission of long wavelength electromagnetic disturbances onto negative energy orbits, when the invariant $B^2-E^2$ becomes negative. In this way, the spinning black hole and its ergosphere not only power the jets but also the ejection disc so as to drive away most of the gas supplied near the Bondi radius. This outflow takes the form of a MHD wind, extending over many decades of radius, with a unidirectional magnetic field, that is collimated by the infalling gas across a magnetopause. This wind, in turn, collimates the relativistic jets and the emission observed from the jet sheath may be associated with a return current. A model for the global flow of mass, angular momentum, energy and current, on scales from the horizon to the Bondi radius, is presented and discussed.
\end{abstract}

\section{Introduction}\label{sec:intro}

Most giant, elliptical galaxies contain massive black holes and some of these create persistent, relativistic jets \citep[see e.g.][for a review]{blandford18}. The jets that are accompanied by intense, ultraviolet radiation and broad permitted emission lines are called radio-loud quasars; those with lower or undetected levels of optical-ultraviolet emission are called radio galaxies. Both types of jet have been associated with discs of accreting gas. In the case of the quasars, the discs are presumed to be radiatively efficient while the radio galaxies are supposed to accrete slowly and to be radiatively inefficient. This paper is concerned primarily with radiatively inefficient accretion in radio galaxies, such as M87.

M87 is, arguably, the best-observed extragalactic radio source. It is associated with a supergiant elliptical galaxy near the centre of the Virgo cluster at a distance of $D=16.8\,{\rm Mpc}$. Two antiparallel jets are observed to propagate out of the nucleus to a projected radius $\sim200\,{\rm as}\sim20\,{\rm kpc}$, and appear to heat the surrounding intracluster medium and moderating the gas inflow and the rate of star formation in the galaxy \citep{russell15}. Radio though $\gamma$-ray telescopes have imaged the jets from 20~MHz frequency to TeV energy \citep{EventHorizonTelescope:2021dvx}. The combined power carried by the two jets is estimated to be $L_{\rm jet}\sim6\times10^{43}\,{\rm erg\,s}^{-1}$ \citep{algaba2021} with an uncertainty $\sim0.5\,{\rm dex}$. The jets appear to be produced by a supermassive black hole whose presence was inferred by stellar dynamical \citep{sargent78,gebhardt11} and gas dynamical \citep{walsh13,nokhrina19} measurements. 

Recent, remarkable observations by the EHT \citep{eht19a} have achieved a resolution (FWHM) of $\sim5m$ and revealed that the source at 230~GHz $\equiv1.3\,{\rm mm}$ comprises a ring with diameter $\sim50\,\mu{\rm as}$ presumably orbiting the black hole and accounting for roughly half of the measured mm, nuclear flux of $\sim1\,{\rm Jy}\equiv L_{\rm ring}\sim10^{41}\,{\rm erg\,s}^{-1}$ \citep{prieto16,algaba2021}. The observed ring has a maximum brightness temperature at position angle (N through E) of $\sim-160^\circ$ of $T_{\rm B}\sim6\,{\rm GK}$ with a secondary maximum at position angle $\sim135^\circ$. For comparison, an upper limit on the bolometric luminosity from within the Bondi radius ($r_{\rm Bondi}\sim250\,{\rm pc}$) is $L_{\rm Bondi}\lesssim10^{42}\,{\rm erg\,s}^{-1}$, much less than $L_{\rm jet}$ \citep{prieto16}.

The black hole mass has been estimated to be $M_H=6.5\times10^9\,{\rm M}_\odot$ \citep{eht19a}. The gravitational radius is then $m\equiv GM_H/c^2\sim3.1\times10^{-4}\,{\rm pc}\equiv3.8\,\mu{\rm as}\equiv \,8.9\,{\rm hr}$. We use these values as units of mass, length and time in what follows and when appropriate. We also suppose that the black hole spin and the angular momentum of the gas supplied are aligned with the counterjet. We adopt, for illustration, a fairly high angular velocity $\Omega_{\rm H}\sim0.3\sim10^{-5}\,{\rm rad\,s}^{-1}\sim0.6$ of the maximal value. This is equivalent to an angular momentum parameter $a\sim0.9$ and corresponds to a rotational energy $\sim0.15\sim2\times10^{63}\,{\rm erg}$, sufficient to keep the circumgalactic gas hot for $\sim10^{12}\,{\rm y}$. The radius of the event horizon is then $r_{\rm H}\sim1.43\sim1.38\times10^{15}\,{\rm cm}$.

 The jet is seen down to a projected radius $\sim60$ and is limb-brightened and steadily collimated on these scales \citep{hada2016}. The receding, counter-jet has an inclination estimated to be $\sim163^\circ$ to the line of sight \citep{walker18} and we also adopt this value for specificity. The jet is not observed directly in the EHT image, but the ring of emission could be associated with the initiation of the jet sheath. 
 
In this paper, we interpret the M87 jets as being 
powered by black hole spin \citep[][and references therein]{eht19a} rather than the release of gravitational energy by conservative accretion through a thick ion-supported torus \citep[e.g.][]{narayan95} or a thin disc \citep[e.g.][]{blandford76,lovelace76,fendt01}. We argue that the black hole not only powers the jet but also transfers sufficient power to the disc to drive away most of the gas that is supplied in the vicinity of the Bondi radius as a MHD wind which, in turn, collimates the jets.  The polarity of the wind magnetic field, equivalently the sense of the current flow, is supposedly constant for an inflow time around the Bondi radius. The standard accretion disc is replaced by an infall at large radius and an ejection disc at small radius. The boundary between the outflow from the ejection disc and the infall is the magnetopause.

In this interpretation, EHT is observing an ergomagnetosphere (an active magnetosphere that allows the transport of angular momentum from the hole to the disc) extending from the horizon to a 
transition radius $\varpi_{\rm em}\sim5$.  This is also where the changeover from a  force-free description to an MHD description occurs.

There are six reasons for pursuing this line of inquiry:\\
$\bullet$ {\it Mass supply.} The nominal gas supply rate at the Bondi radius, $r\sim10^6$, is $\sim10^{25}\,{\rm g\,s}^{-1}$ \citep{allen06,russell15}. This is two orders of magnitude larger than the maximum rate estimated for gas flow onto the black hole \citep{eht21b}, and less than a quarter of the lifetime average. If gas were to flow onto the black hole at the Bondi rate with a conventional efficiency then the disc would be four orders of magnitude more luminous than observed\footnote{The problem is reminiscent of the one posed by the gas in rich clusters of galaxies which were once thought to cause large and invisible cooling flows \citep{fabian94}. It is now believed that these flows are arrested by feedback from the nuclei of the host galaxies.}.\\
$\bullet$ {\it Magnetic field strength.} Presuming that the jet derives from the black hole, its magnetic field strength near the horizon will be $B_{\rm H}\gtrsim100\,{\rm G}$ and the magnetic pressure is $P_{\rm H}\gtrsim1000\,{\rm dyne\,cm}^{-2}$. If the observed emission comes from a thick ring, with volume $\sim 10^{48}\,{\rm cm}^3$, its gas pressure must be at least as large. The energy of the synchrotron-emitting electrons will be $\sim25(B_{\rm ring}/100\,{\rm G})^{-1/2}\,{\rm MeV}$ and their cooling time is $\sim1000(B_{\rm ring}/100\,{\rm G})^{-3/2}\,{\rm s}$. This implies that their partial pressure, averaged over the ring, is $\sim3\times 10^{-5}(B_{\rm ring}/100\,{\rm G})^{-3/2}\,{\rm dyne}\,{\rm cm}^{-2}$, orders of magnitude smaller than the ring pressure, which we presume is dominated by mildly relativistic ions and which should be comparable with the horizon magnetic pressure $\sim1000\,{\rm dyne\,cm}^{-2}$. Again it is surprising that so little of the available free energy is channelled into emitting, suprathermal electrons. This issue is addressed in the EHT analysis by presuming that the electrons can maintain a much lower temperature than the ions with an essentially Maxwellian distribution function and thereby avoid cooling and deflating the torus.\\
$\bullet$ {\it Jet power.} In M87, $L_{\rm jet}\sim600L_{\rm ring}$.
These jets are powered, essentially invisibly, by large scale magnetic field which threads the spinning black hole. This process is highly dissipative but the dissipation takes place invisibly behind the event horizon. However, the jets must also be confined and collimated. If this is effected directly by a trans-sonic, trans-Alfv\'enic and  trans-relativistic shear flow in a thick ion torus, then it is, again, surprising that the torus is so dim.\footnote{Presumably M87 once shone at the Eddington limit and would have been as bright as Vega.}. 
\\
$\bullet$ {\it Continuous jet collimation.} This is observed to be taking place over projected radii $\sim30-10^5m$\footnote{The formation of a long slender jet by flow at the Bondi radius bears some semblance to the formation of a tornado with a width $\sim300\,{\rm m}$ by an atmospheric flow on a scale $\sim10^4$ times larger. The angular momentum comes from large radius; the energy from small radius.}. This is inconsistent with collimation by an electron-ion torus close to the hole. If the collimating agency is static, thermal gas, then its cooling time will be $\sim3(T/10^8\,{\rm K})^{3/2}(P/1\,{\rm dyne\,cm}^{-2})^{-1}\,{\rm yr}$ and it is hard to understand why the confining hot gas is not observed through its X-ray emission. By contrast, if the jet is collimated by the wind, the confining stress will be a combination of magnetic hoop stress and ram pressure, and the shocked wind layer could be responsible for the observed limb-brightening \citep[][]{globus16}. The wind, itself may be effectively invisible. What is clear observationally is that, in M87 and other sources, most of the observed emission is associated with the jet sheath \citep{kim18}. The morphology is what might be expected from a boundary layer where relativistic particles are accelerated, magnetic field is stretched and amplified and thermal plasma from the wind is entrained into the jet flow. Alternatively, as we advocate below, the particle acceleration and emission may be associated with dissipation of a surface electrical current.\\
$\bullet$ {\it Variation in the source.} Although there is, as yet, no compelling evidence for variability in the total intensity over four days \citep{eht19a}, polarization variation is reported on this timescale \citep{eht21a} and this may be surprising, given that the Keplerian period at the radius of the observed ring is roughly a month. A magnetosphere, which can adjust at the speed of light, seems better able to accommodate these changes if they are substantiated.\\
$\bullet$ {\it Polarization of the ring.} The recently reported polarization observations \citep{eht21a} measure significant, variable linear polarization in the brightest part of the ring with azimuthal electric vector and maximum degree $\sim0.15$. The influence of internal and external Faraday rotation complicates the interpretation but the simplest view is that the magnetic field is poloidal and quite strong. This is contrary to what might be expected from a gas-dominated orbiting torus where azimuthal field is expected although simulations \citep{eht21b} can be reconciled with the observations. A further concern is that a larger degree of circular polarization in an ordered magnetic field --- $\sim0.06(B_{\rm ring}/100\,{\rm G})^{1/2}$ at 230 GHz and four times larger at 15 GHz --- is an order of magnitude larger than the upper limits in the nucleus from ALMA, $<0.003$, \citep{goddi21}. A pair plasma should not emit circular polarization but may create it through Faraday conversion \citep[e.g.][]{wardle03}.

This model that we propose has three, inescapable implications.\\
$\bullet$ In order to drive away most of the gas supplied at large radius, requires more black hole power than is being released as gravitational binding energy by the gas. (Just as happened with the sun, we appeal to a larger, central power.) This, in turn, requires that the gas be ejected far from the event horizon so that the disc must transport power outward, as happens in a conventional accretion disc.\\
$\bullet$ There must be sufficient orbiting gas to retain the magnetic flux that extracts energy from the spinning black hole. Confining a $B_{\rm em}\sim100\,{\rm G}$ magnetic field at the transition radius $\varpi_{\rm em}\sim5$ requires the accumulation of $\gtrsim10^{31}\,{\rm g}$ of gas. Note that the accretion rate need not be very large to still allow a relatively massive ring of gas to accumulate over time.\\
$\bullet$ The jet and the ergomagnetosphere are magnetically-dominated and are not usefully described by MHD. They do not exhibit strong dissipation and emission.\\ 
We address each of these points below.

In this paper, we take the view that M87 should be viewed, holistically, on all scales from the event horizon to the surrounding cluster and it is the global transport of conserved quantities --- mass, angular momentum, energy, magnetic flux and current --- that is important for interpreting the EHT observations. 

In Sec.~\ref{sec:disc}, we describe the flow of mass, angular momentum, energy and current, from the infall, where gas flows quasi-spherically to form a geometrically thin ``ejection disc'' out to a radius $\varpi_{\rm ed}\sim 10^5$. Most of the accreted gas  then leaves as  a cold MHD wind, ultimately powered by the ergomagnetosphere which transfers some fraction of the hole rotational energy to the disc. In Sec.~\ref{sec:GREM} we present a general relativistic description of black hole electrodynamics and discuss the jet power. Next, Sec.~\ref{sec:ergomagnetosphere} we discuss our hypothesized ergomagnetosphere, the magnetically dominated region that connects the black hole to the ejection disc and through which torque and angular momentum are transported. In Sec.~\ref{sec:mhdwind}, we examine the dynamics of the MHD wind over several decades of radius. In particular, we describe how  the wind can both collimate the jet and cause the jet sheath to radiate. Our conclusions are briefly summarized in  Sec.~\ref{sec:discuss} and some more general implications are sketched.

In Paper II, we will discuss the ergomagnetosphere and the mechanism to power the MHD wind in more detail. In Paper III, we consider models for emission from the ergomagnetosphere and signatures of its presence that could be revealed by future imaging polarimetry. We also outline some of the ways that the physical mechanisms discussed here could be relevant to a larger class of compact objects.

\begin{figure*}\label{fig:Fig1}
  \begin{minipage}[b]{0.35\textwidth}
       \includegraphics[scale=0.23]{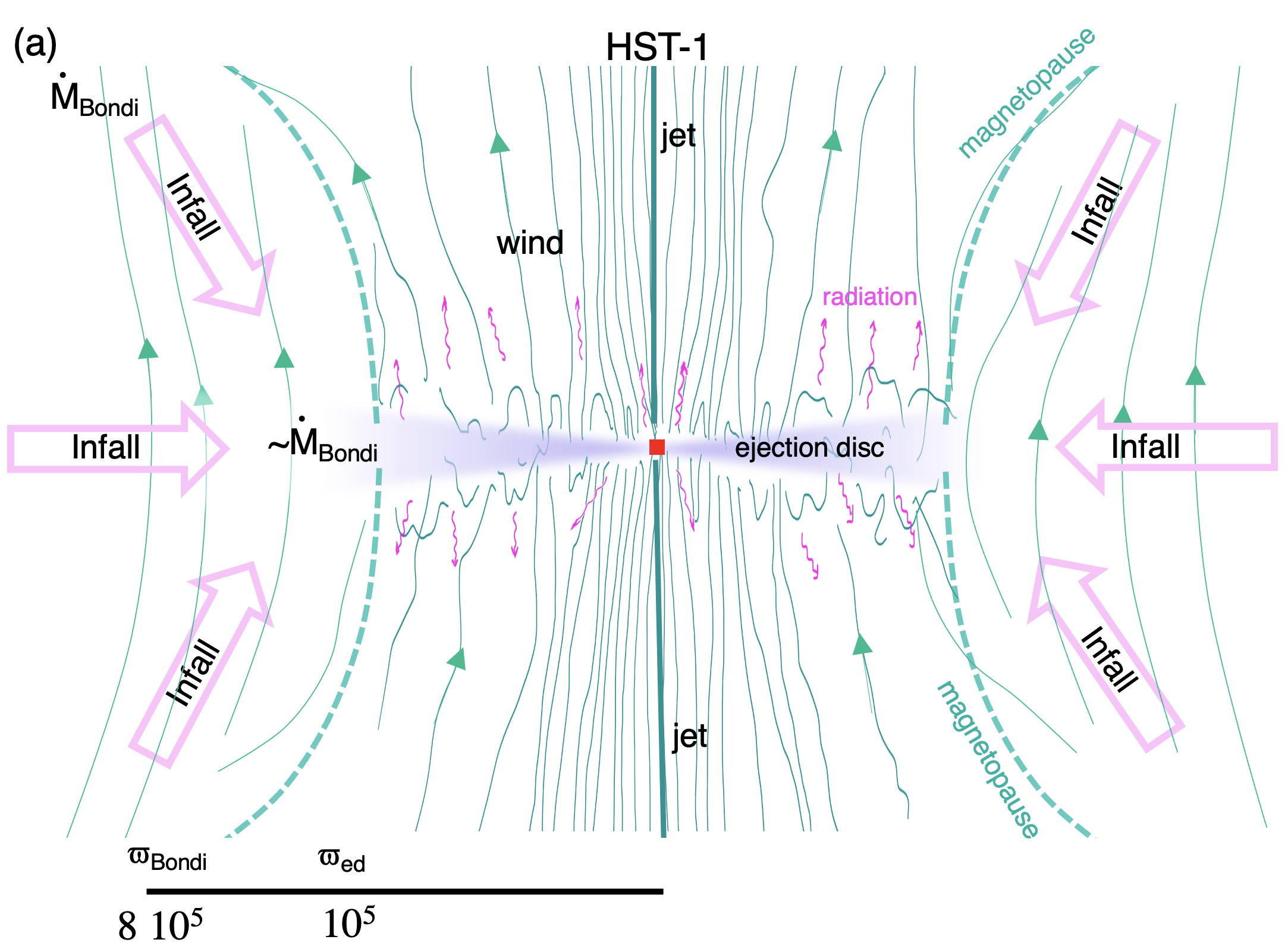}\\
		\includegraphics[scale=0.2]{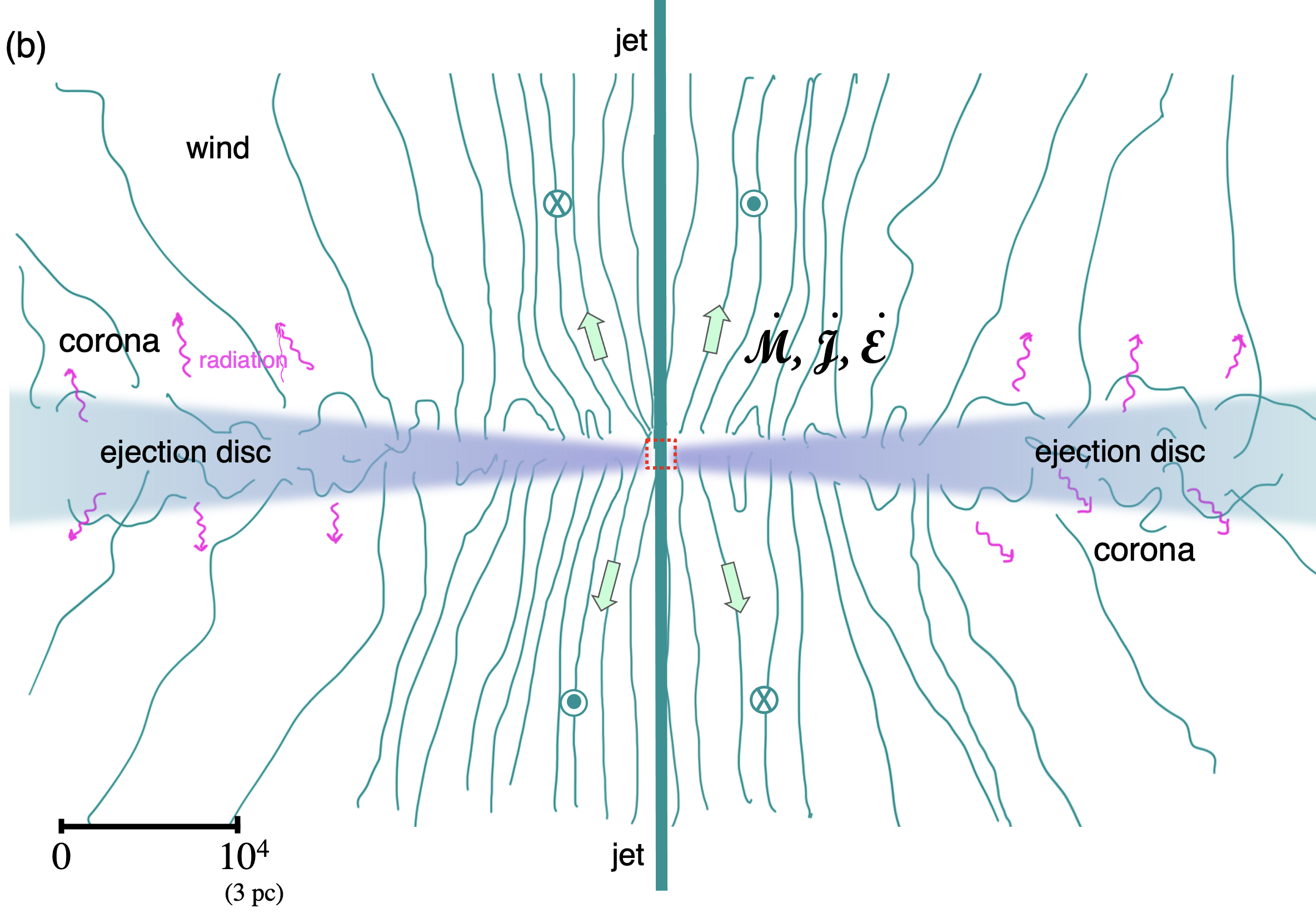}\\\\
		\includegraphics[scale=0.28]{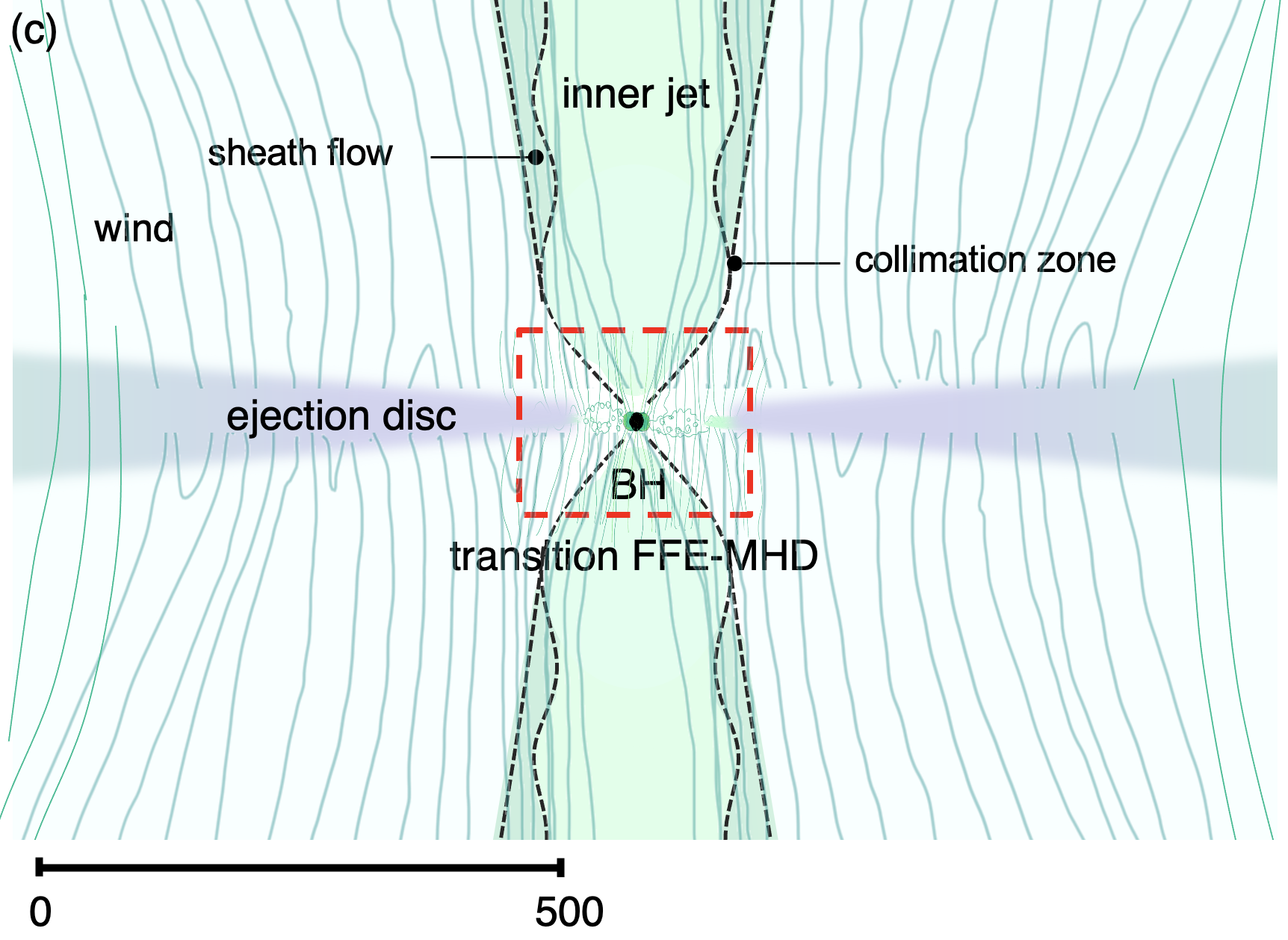}\\\\
				\includegraphics[scale=0.23]{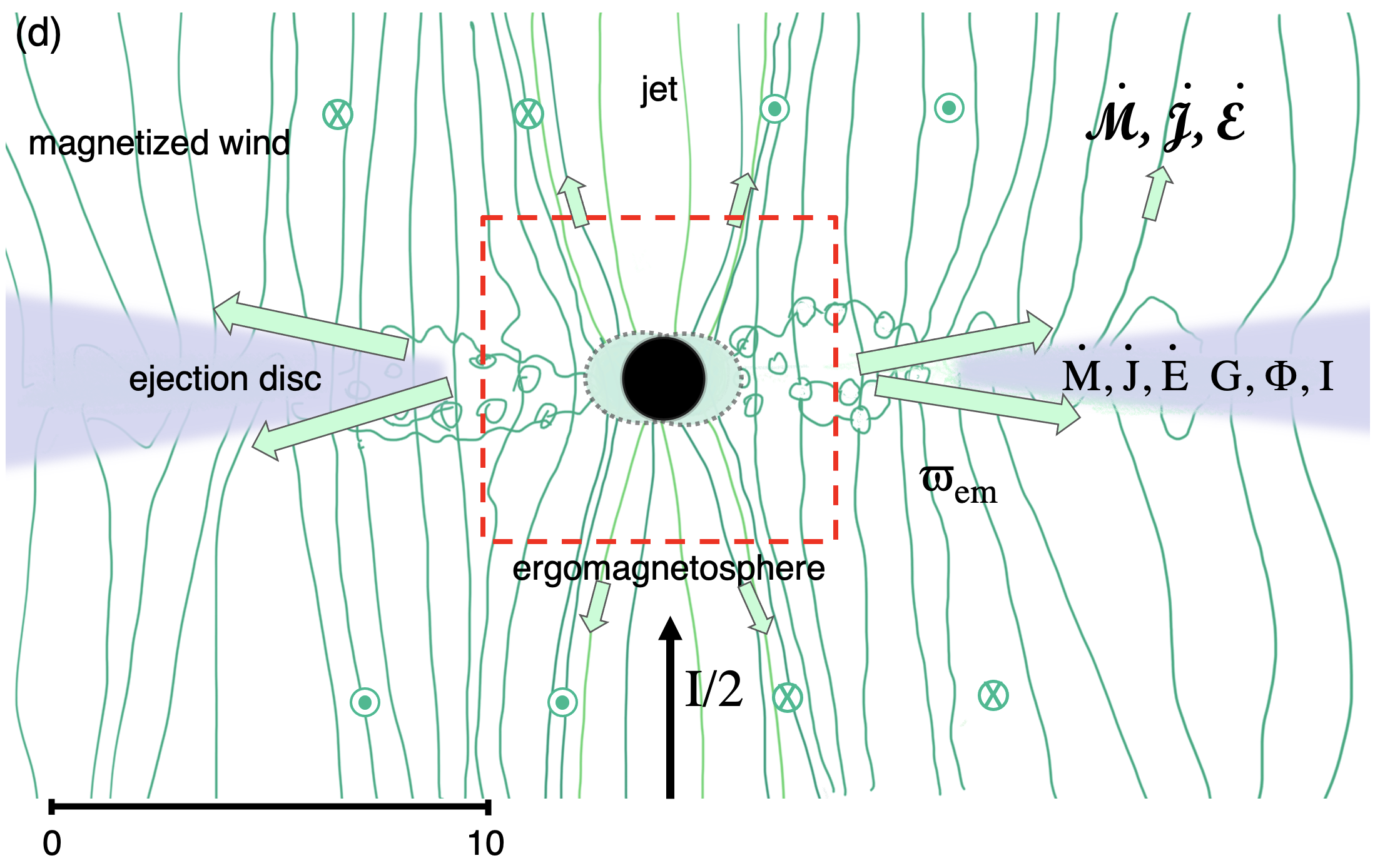}\bigskip
  \end{minipage}\hfill
  \begin{minipage}[b]{0.35\textwidth}
    \caption{Proposed disc dynamics on four radial scales according to the illustrative model.\\ (a) Mass supply around  the Bondi radius, $r_{\rm Bondi}\sim8\times10^5$. The infalling gas is supplied at all latitudes and settles onto an injection disc which transitions to an ejection disc at $\varpi_{\rm ed}\sim10^5$ where the MHD begins. The boundary between the infall and the wind is called the magnetopause passing through $\varpi_{\rm ed}$. This also represents a transition between a flow where MHD is very important to one where other features take over.\\\\\\\\\\\\
    (b) The dense ejection disc, which may contain stars, carries mass and angular momentum inward at rates $\dot M, -\dot J$, with $G$ the torque, and energy outward at a rate $\dot E$. The magnetic flux trapped within radius $\varpi$ is $\Phi$ and current flowing outward in the disc is $I$. The disc is probably covered by a hot corona which launches the MHD wind, carrying off mass, angular momentum and energy at rates $\dot{\cal M},\dot{\cal J},\dot{\cal E}$, and which may radiate away some of the energy dissipated locally in the much denser disc at a rate $\cal L$. \\\\\\\\\\\\\\\
(c) Inner part of the ejection disc for $\varpi<500$. The wind interfaces with the jet across a sheath where much of the current flowing inward along the jets may return and across which mass and linear momentum are transported. The sheath is also  the interface where dynamic and magnetic pressure in the wind collimate the jet. \\\\\\\\\\\\\\\\\\\
(d) Transition from the ejection disc to the ergomagnetosphere within $\varpi\sim15$. The transition occurs at a radius $\varpi_{\rm em}\sim5$. For $\varpi>\varpi_{\rm em}$, mass and angular momentum continue to flow inward. There is likely to be a significant accumulation of mass in a ring which may be unstable. It is unclear how much of the mass remaining at this point is ejected and how much continues inward to the horizon. The outflow changes from a MHD wind to an electromagnetic description across this outflow and the boundary eventually becomes the jet sheath. The ergomagnetosphere is displayed in Fig.~\ref{fig:ergomag}.}\bigskip\bigskip\bigskip\bigskip\bigskip
\end{minipage}\hfill
\end{figure*}

\section{Global Flow of Mass, Angular Momentum, Energy and Current in the Disc}\label{sec:disc}

\subsection{Global Structure}\label{ssec:struc}

We have argued, on observational grounds, that M87 has a large power source near the black hole. It also has a large supply of mass near the Bondi radius while the conspicuous lack of intense, ultraviolet radiation and broad permitted emission lines --- the hallmark of quasars and Seyferts --- suggests that the mass flow through the inner disc is much smaller than this. We have taken this to imply that the power source expels most of the mass in a MHD wind long before it can release a relativistic quantity of gravitational binding energy. In this section, we discuss what this hypothesis implies about the disc behaviour. We identify an infall region, perhaps containing an outer {\it injection disc}, where gas steadily joins the disc flow, and an inner {\it ejection disc} from which there is a mass loss. The ejection disc transitions into the {\it ergomagnetosphere} close to the black hole where the gas is dynamically inconsequential. 

M87's nuclear activity is fueled by gas falling from the surrounding cluster, quasi-spherically towards the black hole Bondi radius, $r_{\rm Bondi}\sim8\times10^5$ \citep{russell2018}, where the escape velocity from the black hole is of order the sound speed and the 1D stellar velocity dispersion, $\sigma\sim400\,{\rm km\,s}^{-1}$. This is about equal to the de-projected radius of HST-1 \citep{algaba21} which may be a recollimation shock that forms in response to a change in the shape of the gravitational potential \citep{levinson2016}. We call the part of the disc where mass is added to the inflow, the {\it infall}.  We suppose the disc at this radius to be, thin, stable and stationary for a flow time, of order a few million years. (The black hole mass and spin should change steadily on timescales up to a thousand times longer.) 

We also assume that the mass flow builds up to a maximum value ${\dot M}_{\rm ed}$ given by the Bondi rate, $\sim10^{25}\,{\rm g\,s}^{-1}$, \citep{allen06,russell15} at a radius $\varpi_{\rm ed}\sim10^5$ 
(see Fig.~\ref{fig:Fig1}). This mass supply is sufficient to produce an AGN disc with luminosity $\gtrsim10^4$ times larger than observed by EHT. The infalling gas also adds angular momentum and energy to the injection disc and it is the former that is more important here\footnote{This is very different from standard theory which is applicable when the angular momentum flow is strictly conservative and a boundary condition of zero torque at the Innermost Stable Circular Orbit, ISCO, determines its value.}.

In addition, we suppose the infalling gas is magnetized with stable polarity, as measured high above the disc\footnote{This view --- that the magnetic field, far above the disc, is simple, ordered and quasi-stable, rather like the quiet solar wind --- is quite distinct from alternative models, where the field is re-generated by disc dynamos on all radial scales and that the sign of the magnetic field in the jet rapidly changes sign.}. This magnetic field exerts a magnetic torque on the disc which contributes heavily to the flow of angular momentum and energy. We estimate that its value is close to the ``Alfv\'en'' value obtained by balancing the ram pressure of the infalling gas with the magnetic stress. This gives $B_{\rm ed}=B(\varpi_{\rm ed})\sim5({\dot M}_{\rm ed}/10^{25}{\rm g\,s}^{-1})^{1/2}(\varpi_{\rm ed}/10^5)^{-5/4}\,{\rm mG}$. The flux threading this part of the disc is $\sim10^4$ times that threading the event horizon of the black hole, needed to power the jets (Sec.~\ref{sec:GREM}). 

In the vicinity of $\varpi_{\rm ed}$, the disc behaves like a traditional accretion disc, radiating away the energy released by the infalling gas and transporting it outward from smaller radii through an internal torque.  At smaller radii, the disc dynamics is dominated by the mass loss and the {\sl outward} flow of energy associated with the {\sl inflowing} gas and an internal torque $G$. There is only modest loss through radiation. This energy flow is supplied by the ergomagnetosphere (Sec.~\ref{sec:ergomagnetosphere}). 

\subsection{Conservation Laws}\label{ssec:claw}
\subsubsection{Mass}\label{sssec:mcon}
In order to make a simple model, we suppose that an axisymmetric, stationary, thin, Keplerian, accretion disc forms within $\varpi_{\rm ed}<\varpi_{\rm Bondi}$. We eschew all discussion of the composition, ionization, density, temperature, cooling, thickness, magnetization, stability, tilt,... of the disc associating it with a time-averaged, {\it disc mass inflow}, ${\dot M}(\varpi)$.  Associated with this inflow, we define a {\it wind mass outflow}, from both disc surfaces
\begin{equation}\label{eq:mcon}
{\dot{\cal M}}=d{\dot M}/d\ln\varpi {\,\,\,\,\,\rm for \,\,\,\,\,} \varpi_{\rm em}< \varpi<\varpi_{\rm ed},   
\end{equation} 
per unit $\ln\varpi$ with ${\dot{\cal M}}(\varpi_{\rm ed})=0$. $\dot{\cal M}<0$ in the infall region, $\dot{\cal M}>0$ in the ejection disc. If we specialise to a power-law variation, $\dot{M}\propto \varpi^n$, \citep[as in][]{blandford99,ostriker95,Casse2000} in the ejection disc, then ${\dot{\cal M}}=n{\dot M}$. This commonly used self-similar scaling describes the mass loss rate in the disc over several decades in radius; however closer to the ergomagnetosphere,  the mass loss rate can be significantly enhanced as there is a torque applied at the inner edge of the disc, as described below. This is where the self-similar scaling breaks down.   

\subsubsection{Angular Momentum}\label{sssec:angcon}
The {\it disc angular momentum outflow}, denoted $\dot J$, contains two terms.  The first is a vertically-, azimuthally-integrated and time-averaged torque $G(\varpi)$. This torque is internal and associated with the $r$--$\phi$ component of the stress tensor. It is probably a consequence of the MagnetoRotational Instability, MRI \citep{balbus98}, so long as the disc is at least partially ionised. The second term represents the angular momentum transported inward by the mass flow. As we are assuming that the disc is thin, we adopt the Keplerian value for the specific angular momentum $\ell_{\rm K}=({\cal G}M_{\rm H}/c)\varpi^{1/2}$, where ${\cal G}$ is the gravitational constant.  Hence ${\dot J}=G-{\dot M}\ell_{\rm K}$. In a traditional accretion disc, it is usually argued that $\dot J\sim0$ except near the inner and outer radii. In an ejection disc, which loses angular momentum through a wind, most of the angular momentum resides at large radius so we anticipate that ${\dot J}<0$. An especially simple and instructive limiting case of a disc - MHD wind flow is obtained by setting $G=0$ (Appendix~\ref{sec:zerot}). 

By contrast, within the ergomagnetosphere, we anticipate that angular momentum flows outward and so ${\dot J}>0$.  
We define the transition radius from the ejection disc to the ergomagnetosphere, $\varpi_{\rm em}$, as being the cylindrical radius, where $\dot J$ changes sign.

Within the ejection disc, we define the {\it wind angular momentum outflow} per $\ln\varpi$, by $\dot{\cal J}={\dot{\cal M}}\ell$, where $\ell$ is the total specific angular momentum in the wind, which we assume to be conserved along along a flux surface under perfect MHD (Sec.~\ref{sec:mhdwind}). $\ell$ is the sum of a gas component, $\ell_{\rm g}$, and a magnetic component $\ell_{\rm m}$. Conservation of angular momentum then implies that 
\begin{equation}\label{eq:jcon}
\frac{d{\dot J}}{d\ln\varpi}=\frac{d}{d\ln\varpi}(G-\dot M\ell_{\rm K})=-{\dot{\cal J}}=-{\dot{\cal M}}\ell=-\dot{\cal M}(\ell_{\rm g}+\ell_{\rm m}),
\end{equation}
generalizing \cite{blandford99,begelman12}. 

If we assume $\varpi_{\rm em}<<\varpi<<\varpi_{\rm ed}$, and adopt a power form for the disc mass inflow, and setting $\ell_{\rm g}$ to $\ell_{\rm K}$ at the disc, Eq.~(\ref{eq:jcon}) becomes
\begin{equation}\label{eq:Gpowerlaw}
G=\left(\ell_{\rm K}-\frac{2n\ell}{2n+1}\right){\dot M}\,.
\end{equation}

\subsubsection{Energy}\label{sssec:encon}
In a similar fashion, the {\it disc energy outflow}, $\dot E$ combines the rate of work done by the torque, $G\Omega_{\rm K}$ and the transport outward of the orbital energy, $e_{\rm K}$. Therefore, ${\dot E}=(G+\frac12{\dot M}\ell_{\rm K})\Omega_{\rm K}$, where $\Omega_{\rm K}=(c^3/{\cal G}M_{\rm H})\varpi^{-3/2}$. We define the rate at which the outflow carries energy away, per $\ln\varpi$, from both disc surfaces by ${\dot{\cal E}}={\dot{\cal M}}e+{\cal L}$, where $e$ is the specific energy in the outflow, which is also conserved under perfect MHD, and $\cal L$ is the power radiated as photons per $\ln\varpi$\footnote{For a radiation-driven wind, $\dot{\cal E}$ and $\cal L$ are coupled but the wind kinetic energy is typically a fraction $O(fv/c)$ of the luminosity, where the fraction $f\sim (v/c)$ per line for line-driving and $\sim\tau_{\rm T}$ for Thomson scattering. This coupling is not important for our model.}. Conservation of energy then implies that 
\begin{equation}\label{eq:econ}
\frac{d{\dot E}}{d\ln\varpi}=\frac d{d\ln\varpi}[(G+\frac12{\dot M}\ell_{\rm K})\Omega_{\rm K}]=-{\dot{\cal E}}=-{\dot{\cal M}}e-{\cal L}.
\end{equation}
where $e$ is the total specific energy of the wind which is also conserved along a flux surface under MHD. At the disc, $e$ has three contributions, an orbital energy $e_{\rm K}=-\Omega_{\rm K}\ell_{\rm K}/2$, a magnetic term $\Omega_{\rm K}\ell_{\rm m}$ and a starting enthalpy $h_0$ (see Sec.~\ref{ssec:launchw}).

Again, if we adopt a power law mass inflow, then the requirement that the outflow be able to escape, $e>0$, becomes $n<1$. A more stringent condition must be satisfied if radiative loss is significant or the wind has significant kinetic energy at infinity.

\subsection{Magnetic Field and Electrical Current}\label{ssec:mhdio}
We have proposed that mass, angular momentum and energy are removed from the disc by a MHD wind. We defer discussion of the wind itself till Sect.~\ref{sec:mhdwind}. Here we just consider the implications for the disc.

The magnetic field is conventionally described in terms of its poloidal and toroidal components, $B_p$, $B_\phi$. We replace these with two equivalent quantities, the large-scale, time and azimuth averaged magnetic flux $\Phi(\varpi)$ threading a circle of radius $\varpi$, and an electric current $I(\varpi)$ flowing radially outward in the disc across this circle. The smaller scale field within the disc and its corona is responsible for the torque, $G$.

We adopt the convention that the angular velocity of the black hole and the disc defines the $+z$ axis and that 
\begin{equation}\label{eq:magfield}
B_z=\frac{1}{2\pi\varpi^2}\frac{d\Phi}{d\ln\varpi}>0;\,B_\phi=-\frac I\varpi<0
\end{equation}
We evaluate $B_\phi$ above the corona in the upper hemisphere.

Using the standard theory of MHD winds, the magnetic part of the angular momentum flow in the wind should be 
\begin{equation}\label{eq:jmdot}
\dot{\cal J}_{\rm m}={\dot{\cal M}}\ell_{\rm m}=-\varpi^3B_zB_\phi=\frac I{2\pi}\frac{d\Phi}{d\ln\varpi}\,.
\end{equation}
The corresponding magnetic contribution to the energy flow away from the disc is ${\dot{\cal E}}_m=\Omega_{\rm K}{\dot{\cal J}}_m$.

We need to make one further assumption to define a model and choose to suppose that the ratio $B_\phi/B_z$ maintains a constant value at the disc surface. An estimate based upon the discussion of Sec.~\ref{sec:mhdwind} suggests $B_\phi/B_z=k\sim-0.1$. The actual ratio depends upon the shape of the flux surfaces and the boundary conditions and is unlikely to be constant in practice. On this basis, we can use Eq.~(\ref{eq:jmdot}) to solve for the magnetic field variation. The answers are almost surely underestimates as most of the disc surface may be threaded by closed field lines, but they do demonstrate that quite modest values of the magnetic field strength suffice to eject almost all of the mass that is supplied. 

\subsection{Dissipation}
Equations~(\ref{eq:mcon}-\ref{eq:econ}) can then be combined to give
\begin{equation}\label{eq:diss}
-G\frac{d\Omega_{\rm K}}{d\ln\varpi}=\frac32G\Omega_{\rm K}={\dot{\cal M}}h_0+{\cal L}.
\end{equation}
This identity is simply interpreted as equating the radiative loss plus the thermal heating of the outflow with the dissipation that inevitably arises when a differentially-rotating shear flow develops a shear stress. Note that this is independent of the fraction of the angular momentum carried off magnetically. It captures the essence of how an ejection disc should operate. In particular it shows that the torque need only be large enough for its dissipation to heat coronal gas and elevate it to sufficient altitude that it can be flung out centrifugally by the magnetic field \citep{blandford82}. Eq.~(\ref{eq:diss}) demonstrates another important property of the solutions to these equations, that for a given dissipation rate, the thermal heating is degenerate with the surface radiation. 

\subsection{Illustrative solution}\label{ssec:illustris}

At this stage, it is helpful to set down an illustrative solution for the flow in the ejection disc where we make simplifying assumptions to explain some of the constraints that should be satisfied by a more careful treatment and to demonstrate that solutions can be found that satisfy our interpretation of existing observations.

Our starting point is to adopt the estimated Bondi accretion rate $\sim10^{25}$~g~s$^{-1}$ as the maximum gas flow at the outer radius of the ejection disc, which we choose as $\varpi_{\rm ed}\sim10^5\equiv10^{20}\,{\rm cm}$.  For the inner/transition radius, we adopt $\varpi_{\rm em}\sim5\equiv5\times10^{15}\,{\rm cm}$. Within these limits, we assume a power-law for the mass inflow rate with $n=0.8$ and derive a simple self-similar solution. This allows the wind to escape and limits $\ell$ to $\lesssim1.7\ell_K$, if $G>0$ (Eq.~(\ref{eq:Gpowerlaw})). We make the intermediate choice $\ell=1.4 \ell_K$ so that $G\approx0.14\dot{M}\ell_K$. We also adopt $k\sim0.1$ (Sec.~\ref{ssec:mhdio}). The dissipation associated with the internal torque needs to be partitioned between heating the coronal gas ${\dot{\cal M}}h_0$ and radiating at rate per $\ln\varpi$ of $\cal L$ and we make the arbitrary choice that they are equal.

\begin{figure*}
\begin{center}
\centering
\includegraphics[width=\textwidth]{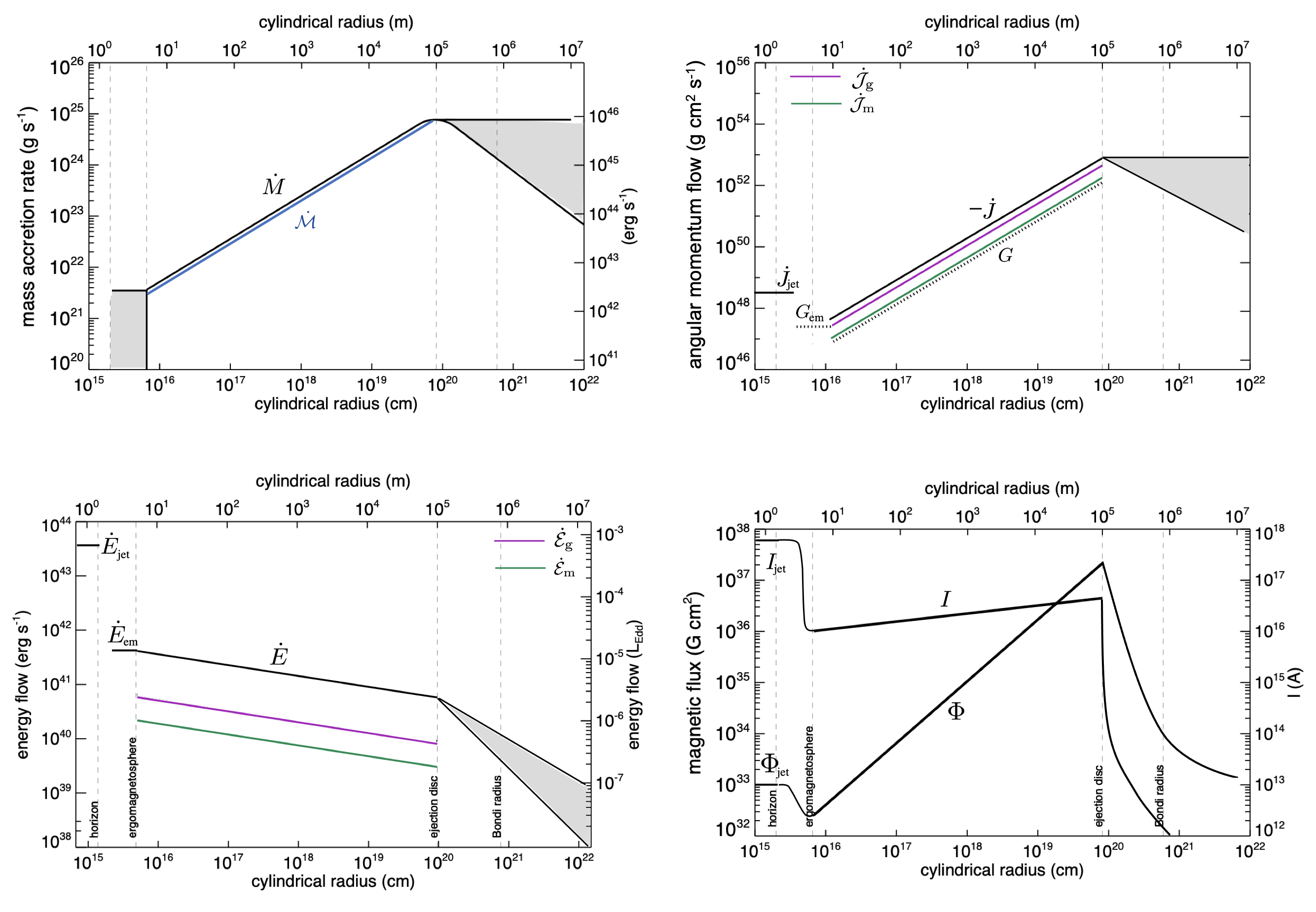}
\caption{Global flow of mass, angular momentum, energy and current on scales from the event horizon to the Bondi radius, adopting our illustrative model, Eq.~(\ref{eq:illsol}). a) The mass flowing through the disc, $\dot M$, is similar to the mass leaving the disc per $\ln\varpi$, $\dot{\cal M}$. In the infall region beyond $\varpi_{\rm ed}$, there may be either an accretion disc or an injection disc, as shown schematically in the shaded area. Within $\varpi_{\rm em}$, the gas flow can range from vanishing small to that found in the ejection disc. b) Angular momentum flows inward through the ejection disc, carried mostly by the matter and opposed by the torque $G$. Angular momentum also leaves the disc at rates $\dot{\cal J}_{\rm g}, \dot{\cal J}_{\rm m}$ for the gas and magnetic parts respectively. Within the ergomagnetosphere, electromagnetic torque, $G_{\rm em}$,  transports angular momentum outward through a variety of mechanisms. Note that the angular momentum flow in the jet, some of which may derive from the ergomagnetosphere, is significantly larger than that flowing inward in the disc and only a fraction of this need be tapped to change the sign of $\dot J$. c) By contrast, the energy flow is outward in both the ergomagnetosphere and the ejection disc and is roughly $\sim0.01$ times the jet power. The disc radiative luminosity $\cal L$, which is assumed to equal the coronal heating, is relatively small. In this simple example, only a fraction $\sim0.01$ of the jet power, ${\dot E}_{\rm jet}$, is needed to drive away all of the gas supplied. In a more realistic description, the fraction may be  higher, $\sim0.1$. d) The magnetic flux, $\Phi$, grows rapidly with $\varpi$ in the disc and only a very small fraction of what is available at large radius is needed to power the jets. In this model, the magnetic field strength declines rapidly through the transition radius and then declines more slowly within the disc. The current $I$ is large along the jet and much smaller in the disc. This is consistent with there being a large current flowing out from $\varpi_{\rm em}$ into the jet sheath and contributing to the observed, limb-brightened radio emission.} 
\label{fig:fig2}
\end{center}
\end{figure*}

We can then solve the preceding equations to evaluate the flow of mass, angular momentum, energy and current\footnote{Astronomers mostly prefer cgs units but are also more familiar with SI units for electrical quantities. Care must be used when they appear in the same equation. In particular, it should be recalled that the cgs unit of current, the Biot (Bi), equals 10~A.} in the ejection disc on the basis of these assumptions. The results are displayed in Fig.~\ref{fig:fig2}.
\begin{align}\label{eq:illsol}
\{\dot M,\dot{\cal M}\} &=\{10,8\}\times10^{20}\varpi_{15}^{0.8}\,{\rm g\,s}^{-1}\nonumber\\
\{\dot J,G,\dot{\cal J}_{\rm g},\dot{\cal J}_{\rm m}\}&=\{-30,4,15,6\}\times10^{45}\varpi_{15}^{1.3}\,{\rm g\,cm}^2\,{\rm s}^{-2}\nonumber\\
\{\dot E,{\cal L},\dot{\cal E}_{\rm g},\dot{\cal E}_{\rm m}\}&=\{60,9,8,3\}\times10^{40}\varpi_{15}^{-0.2}\,{\rm erg\,s}^{-1}\nonumber\\
\{B_{\rm z},B_\phi\}&=\{7,0.7\}\,\varpi_{15}^{-0.85}\,{\rm G}\nonumber\\
\Phi&=4\times10^{31}\varpi_{15}^{1.15}\,{\rm G\,cm}^2\nonumber\\
I&=8\times10^{15}\varpi_{15}^{0.15}\,{\rm A}
\end{align}

Several comments can be made at this stage. The assumptions that we have made are chosen to optimise the removal of the mass supplied at $\varpi_{\rm ed}$ throughout the disc. The angular momentum flows inward; the energy flows outward. The disc current increases with radius resulting in an inward magnetic force on the wind. It is found that a comparatively modest magnetic field suffices in the MHD wind and within the disc and may be necessary. If too much angular momentum is released per unit mass then little mass will be lost and more mass will be supplied at $\varpi_{\rm em}$. If the internal torque $G$ is larger, as in a conventional disc, there will be more dissipation and radiation than is observed. There are many reasons why we expect the wind to be less efficient, permitting a larger value of the wind magnetic field and facilitating the confinement of more magnetic flux around the spinning black hole. In addition, the assumption of a power law, self-similar solution precludes discussion of the inner and outer radii which can seriously affect the actual solution. Despite this, the solution does provide an existence proof for mass ejection driven by a central power source.

\section{Black Hole Force-free Electrodynamics}\label{sec:GREM}
In this section, we reconsider the powering of the jets by a spinning black hole. We focus attention on the magnetic flux that threads the event horizon. We discuss how the confinement of the magnetic flux and remote electromagnetic boundary conditions determine the power extracted.

\subsection{Basic Equations and Approaches}\label{ssec:basequ}
So far our discussion has been mostly conducted without appeal to general relativity. The properties of black holes and ergomagnetospheres are peculiarly relativistic. In order to understand the approximations we are making in this paper, we need to outline some different approaches to black hole electrodynamics. 

We start with a vector potential $\vec A$, over the 4D spacetime constrained by the Lorenz gauge, ${\vec\nabla}\cdot{\vec A}=0$\footnote{We follow the approach and index-free notation of \cite{thorne17}, where $\vec\nabla$ is a gradient that operate on a scalar, vector or tensor, adding the common conventions that adjacent slots/indices are involved in dot products and : denotes two successive dot products.}. We next define the electromagnetic field tensor ${\ten F}$ as twice the antisymmetric part of ${\vec\nabla}\vec A$. Electric and magnetic field can be read off in a local Lorentz frame from this in the normal manner. The field tensor is used in three different ways. 

Firstly for a particle with mass $m$ velocity $\vec U$ and charge $q$, the electromagnetic force is $q {\ten F}\cdot{\vec U}$. This is orthogonal to $\vec U$, as it must be, and can be used to update the position of the particle. This is the basis of a ``PIC'' code \cite[e. g.][]{parfrey19}. The current density 
\begin{equation}\label{eq:vecj}
\vec j={\vec\nabla}\cdot{\ten F}/4\pi
\end{equation}
is estimated by summing many particles.

The second use of the field tensor is fluid-based. It supposes that the plasma can be described locally by its centre of momentum frame velocity $\vec u$. The justification of this approach when the plasma is collisionless is not straightforward but generally follows from taking  moments of kinetic equations under various approximations and imposing closure.  In ``perfect'' MHD it is supposed that the electric field vanishes in this center of momentum frame. This implies that ${\ten F}\cdot{\vec u}=0$. This is commonly referred to as a statement of Ohm's law in the infinite conductivity limit. It can be replaced by a finite conductivity scalar Ohm's law or more realistic generalizations that involve Hall currents and entropy production \citep[e.g.,][]{koide2009}. In perfect MHD, the field lines are Lie-transported by the fluid in a local Lorentz frame and the evolution of the velocity field follows from setting the divergence of the combined fluid and electromagnetic stress-energy tensor to zero. The plasma pressure may be anisotropic, isotropic or ignored. Relativistic MHD is the basis of most simulations of the M87 ring \citep[e.g.][]{eht19b} because it is envisaged that the plasma pressure and inertia dominate the magnetic stress in the ring.

The third approach is force-free electrodynamics. Here it is supposed that the density of current-carrying charged particles is large enough to keep the electric field components along the mean magnetic field and the contribution of the plasma to the stress-energy tensor at an ignorably low level. Under these conditions, the electromagnetic field will satisfy the relativistic force-free condition 
\begin{equation}\label{eq:ffc}
{\ten F}\cdot{\vec j}={\vec\nabla}\cdot{\ten T}=0. 
\end{equation}
where $\ten T$ is the stress energy tensor.
\begin{equation}\label{eq:TEM}
{\ten T}=-\frac1{4\pi}({\ten F}\cdot{\ten F}-\frac14{\ten F}:{\ten F}{\ten g}),
\end{equation}
and ${\ten g}$ is the metric tensor. In other words, the contribution to the total stress energy tensor from the plasma and the radiation can be ignored. 

In this approach, the electrical current is regarded as a continuous flow and the microphysical details are unimportant for addressing the large scale transport of 4-momentum\footnote{This is much the way an oceanographer works, for example. A transition to turbulence is handled using the same Navier-Stokes equations as are used for streamline flow and the character of the water molecules is adequately captured by the viscosity which is usually unimportant. Interestingly, Maxwell in the 1860's treated current in this manner, despite his acquaintance with Faraday's laws of electrolysis and his pioneering work on the kinetic theory of gases. It was not until around 1890 that the electromagnetic force on a charged particle was widely discussed, leading eventually to atomic, nuclear, particle and plasma physics.}. Of course, the microphysics can be crucial if we are trying to calculate the emissivity, especially in $\gamma$-rays, but that is not our primary concern here. 

There are two Lorentz  scalars associated with $\ten F$. The first of these evaluates to
\begin{equation}\label{eq:lorinv}
{\cal I}=\frac12{\ten F}:{\ten F}=B^2-E^2,
\end{equation}
in a local Lorentz frame is usually positive but can evolve to become negative as we discuss below (Sec.~\ref{ssec:governor}). The second invariant evaluates to ${\bf E}\cdot{\bf B}$ and vanishes under force-free conditions.

\subsection{Justification of the Force-free Approximation}\label{sssec:FFEM}
Although there have been many, instructive investigations of classical electromagnetism in a Kerr spacetime without currents and charges \citep[e.g.][]{wald74}, these are unlikely to be relevant to sources like M87. This is because when power is extracted electromagnetically from a spinning black hole, the potential difference across the horizon, $\Delta V$ will satisfy $L_{\rm jet}\sim\Delta V^2/Z_0$, where $Z_0$ is the impedance of free space. So $\Delta V$ could be as large as $\sim30\,{\rm EV}$. This is over $10^{13}$ times the minimum potential difference needed to create an electron positron pair $\sim1\,{\rm MV}$. This is also roughly the ratio of the size of the horizon to the gyro radius of a mildly relativistic electron and so this can never be represented faithfully in a PIC simulation. In addition, it is the ratio of the magnetic energy density to the minimum, mildly relativistic, electron-positron energy density needed to sustain the current. 

The particular QED processes at work have been much discussed and debated \citep{levinson18,yuan19,parfrey19}. Under typical astrophysical conditions in a galactic nucleus, we may need $\sim1\,{\rm TV}$ of potential difference across a ``gap'' to create $\gamma$-rays energetic enough to pair-produce on optical photons \citep{levinson2011,Chen2020}. Other mechanisms may operate faster. Charge creation should happen rapidly and locally, shorting out small components of electric field along the direction of the magnetic field. The perpendicular electric field can be removed by transforming into a frame moving with speed ${\bf E}\times{\bf B}/B^2$, presuming $E<B$. Furthermore the multiplicity, the ratio of the actual pair density to the minimum needed electrodynamically, could be as high as a million, if we are guided by pulsars. However, this still leaves plenty of room for the plasma around a black hole to supply all the charge and current required while being dynamically insignificant. This is the force-free limit.

It may not even necessary to create {\it any} pairs around the black hole. This is because the event horizon, is only a moderate electrical conductor, unlike a neutron star which line-ties the magnetic field. This allows the field lines to interchange throughout the magnetosphere carrying trapped particles with them all the way from the injection disc where there is an abundance of plasma. This transport may be sufficient to resupply the very few charges needed to carry electrical current along the jet and through the horizon. In this case, the current comprises electrons and protons, not electrons and positrons, a distinction that has observational implications, as we discuss in Paper III.

We therefore argue that in the vicinity of the black hole, specifically at the base of the jets and also in the ergomagnetosphere, that the electromagnetic stress overwhelms the plasma stress and the force-free formalism is the best one to use. However, this formalism must fail at high altitude when there will be a significant entrainment of matter into the jets. It will also fail in the vicinity of $\varpi_{\rm em}$ in the equatorial plane where the disc becomes gas-dominated. Provided that $\varpi_{\rm em}$ is not too small, the non-relativistic MHD approach we have used in Sec.~\ref{sec:disc} should then suffice.

\subsection{Stationary Axisymmetric Force-Free Electrodynamics in the Kerr spacetime}\label{ssec:SAE}
\subsubsection{Flux, Potential, Angular Velocity and Current}\label{sssec:FVWI}
So far, our discussion has been quite general. We now restrict our attention to the idealization of a stationary, axisymmetric electromagnetic field in a \cite*{kerr63} spacetime with temporal and azimuthal Killing vectors, $\vec \xi$ and $\vec\eta$ respectively. More precisely, we suppose that the average electromagnetic field shares these symmetries. We summarize the results of more didactic treatments \citep[e.g.][]{thorne86}. We consider non-stationarity and non-axisymmetry in Sec.~ \ref{sec:ergomagnetosphere} and Paper II.

We introduce the scalar $\Phi=2\pi{\vec A}\cdot{\vec\eta}$ as the magnetic flux contained within stationary and axisymmetric, nested flux surfaces. Its value is independent of the coordinate system chosen. We assume that the flux surfaces are similar in the northern and southern hemispheres. We also define the electrical potential, $V=-{\vec A}\cdot{\vec\xi}$, another scalar. $V$ is set to zero on the axis of symmetry. 

Using the force-free condition, Eq.~(\ref{eq:ffc}), we deduce immediately that the flux surfaces are equipotentials, $V=V(\Phi)$, This is an expression of the local, Lorentz invariant relation ${\bf E}\cdot{\bf B}=0$.  Furthermore, the derivative $2\pi dV/d\Phi$ can be identified with the electromagnetic angular velocity, henceforth $\Omega$ which should generally lie between 0 and $\Omega_{\rm H}$ with a value determined by the imperfect conductivity associated with the horizon and the properties of the outflow, as we discuss further in Sec.~\ref{sssec:causal}.

We next consider the 4D current vector $\vec j$. Its 4D divergence vanishes in general. Imposing stationarity and axisymmetry allows us to extract the $r,\theta$ components as a 2D poloidal current vector with vanishing 2D divergence. Eq.~(\ref{eq:ffc}) then implies allows us to infer that the current flows along the flux surfaces. We can then define a scalar, poloidal current $I(\Phi)$ flowing into the hole within a given flux surface $\Phi$, summing over  both hemispheres (cf. Sect~\ref{ssec:mhdio})\footnote{Having more negative than positive charges cross the horizon produces an outward current.}.

\subsubsection{Angular Momentum and Energy Flows}\label{sssec:AME}
Combining the vanishing divergence of the electromagnetic stress-energy tensor, Eq.~(\ref{eq:TEM}), with the Killing equation implies that the two 4-vectors $ {\ten T}\cdot{\vec\xi}$, ${\ten T}\cdot{\vec\eta}$, associated with energy and angular momentum, are conserved in the sense that their 4D divergences vanish. Again, under stationarity and axisymmetry, we can define 2D poloidal energy and angular momentum flux vectors lying in the magnetic flux surfaces. (This flow of energy along the flux surfaces is a consequence of its being a Poynting flux plus the condition ${\bf E}\cdot{\bf B}=0$.) We can then define the electromagnetic power ${\dot E}(\Phi)$ and the rate of flow of angular momentum, ${\dot J}(\Phi)$, interior to a given flux surface, summing over both hemispheres.

In Secs~\ref{sssec:angcon}, \ref{sssec:encon}, we introduced corresponding differential quantities, $\dot{\cal E}$, $\dot{\cal J}$ to represent the flux densities of energy and angular momentum per unit $\ln\varpi$ at the magnetic footpoints in the disc. In this section, it is more instructive to consider these flows as per unit magnetic flux, $\Phi$. We denote these quantities, $\dot{\mathbb E}=d{\dot E}/d\Phi$, $\dot{\mathbb J}=d{\dot J}/d\Phi$, respectively. These are also scalars conserved along flux surfaces, like $V, \Omega, I$. In general, we find that
\begin{equation}\label{eq:powerflux}
{\dot{\mathbb E}}=\Omega {\dot{\mathbb J}}=\frac{\Omega }{2\pi}I=I\frac{dV}{d\Phi},
\end{equation}
{\it cf} Eq.~(\ref{eq:jmdot}). In words, the angular momentum carried off per unit magnetic flux is simply given by $I/2\pi$.

The identification of the increment of energy flow with $IdV$ can be interpreted using a simple circuit with a continuous current $I$ flowing inward along one flux surface, crossing onto the next flux surface within the horizon, flowing outward along this second surface and then returning to the first flux surface in a jet in a region called the ``load''. The relation between the increments of power and torque is also generic as it is the rate of performing work by the electromagnetic torque/angular momentum flow. 

In summary, under stationarity and axisymmetry, $V,\Omega,I,{\dot{\mathbb E}},{\dot{\mathbb J}}$ are scalars, constant on surfaces of constant $\Phi$.

\subsubsection{Coordinates and Frames}\label{sssec:kerr}
In order to exhibit solutions of the force-free equations, we need to introduce a coordinate system. The existence of Killing vectors suggests introducing a time coordinate, $t$, such that ${\vec\xi}=\partial_t$ and an azimuthal coordinate, $\phi$, such that ${\vec\eta}=\partial_\phi$. We add a radial coordinate $r$ and a polar angular coordinate, $\theta$ chosen so that the event horizon is at $r=r_{\rm H}$. This leads to the definition of \cite*{boyer67} coordinates, $x^{\rm B\alpha}\equiv\{t,r,\theta,\phi\}$ with corresponding coordinate basis vectors ${\bf e}^{\rm B}_\alpha=\{\partial_t,\partial_r,\partial_\theta,\partial_\phi\}$. This basis allows us to define vectors and tensors in a tangent space at a point. We call this the B frame. The time coordinate, $t$, measures ``time at infinity'' which advances more slowly than local time near the event horizon. The line element is
\begin{equation}\label{eq:boylin}
ds^2=-\alpha^2dt^2+\rho^2(dr^2/\Delta+d\theta^2)+\varpi^2(d\phi-\omega dt)^2,
\end{equation}
where the lapse $\alpha=\rho\Delta^{1/2}/\Sigma$, the Zero Angular Momentum Observer, ZAMO, angular velocity is $\omega=2ar/\Sigma^2$ and the cylindrical radius is $\varpi=\Sigma\sin\theta/\rho=\Delta^{1/2}\sin\theta/\alpha$. Lengths and times are measured in terms of the gravitational radius and the spin parameter $a$ satisfies $-1<a<1$. In turn, $\rho^2=r^2+a^2\cos^2\theta$, $\Delta=r^2-2r+a^2$, and $\Sigma^2=(r^2+a^2)^2-\Delta a^2\sin^2\theta$. The determinant associated with the metric tensor is then $g=-\rho^4\sin^2\theta$. The event horizon is at $r=r_{\rm H}=1+(1-a^2)^{1/2}$ and the ergosphere extends out to $r=r_{\rm E}=1+(1-a^2\cos^2\theta)^{1/2}$. Inspecting the metric tensor, Eq.~(\ref{eq:boylin}), we see that the 4D spacetime separates into $r-\theta$ and $t-\phi$ subspaces. 

The ZAMO frame (hereafter Z frame), is, by definition, the local frame with zero angular momentum, orbiting the hole at fixed $r,\theta$ with fixed angular velocity $\omega$.
It is convenient to define another local Lorentz "M" frame, the Magnetic frame, orbiting the hole at fixed $r,\theta$ with fixed electromagnetic angular velocity $\Omega$. The transformation matrices between the different frames B, Z, and M are given in  Appendix~\ref{sec:cotrans}.

The two light surfaces are defined by
\begin{equation}
    \frac{\varpi(\Omega-\omega)}{\alpha }=\pm 1
\end{equation}
which implies $\Omega_{\rm min}<\Omega<\Omega_{\rm max}$ with $\Omega_{\rm min}=\omega-\alpha /\varpi$, $\Omega_{\rm max}=\omega+\alpha /\varpi$.

\subsubsection{Electromagnetic Field}\label{sssec:emfld}
Consider first, the field tensor in the B frame.
\begin{equation}\label{eq:FB}
F^{\rm B}_{\alpha\beta}=
\begin{pmatrix}
0&\frac{\Omega\Phi_{,r}}{2\pi}&\frac{\Omega\Phi_{,\theta}}{2\pi}&0\\
-\frac{\Omega\Phi_{,r}}{2\pi}&0&-\frac{I\rho^2}{\Delta\sin\theta}&\frac{\Phi_{,r}}{2\pi}\\
-\frac{\Omega\Phi_{,\theta}}{2\pi}&\frac{I\rho^2}{\Delta\sin\theta}&0&\frac{\Phi_{,\theta}}{2\pi}\\
0&-\frac{\Phi_{,r}}{2\pi}&-\frac{\Phi_{,\theta}}{2\pi}&0
\end{pmatrix}.
\end{equation} 
This 4D, second rank, antisymmetric tensor is well-defined and satisfies the Einstein-Maxwell equations in this frame, even within the ergosphere. 

It is instructive to consider the electric and magnetic field in the M frame of a physical observer with $\Omega_{\rm min}<\Omega<\Omega_{\rm ed}$. Using Eq.~(\ref{eq:transBO}), We confirm that the electric field vanishes. Define the 2D gradient operator ${\tilde\nabla}\equiv\rho^{-1}\{\Delta^{1/2}\partial_r,\partial_\theta\}$ in the $r-\theta$ plane. 
The regular, 3D magnetic field, as measured by an observer in the M frame can then be read off as
\begin{equation}\label{eq:Mmagf}
{\bf B}^{\rm M}=-\left(\frac\beta{2\pi\alpha\varpi}\right){\vec e}_{\hat\phi}\times{\tilde\nabla}\Phi-\frac I{\alpha\varpi}{\vec e}_{\hat\phi}\,,
\end{equation}
where $\beta=(\alpha^2-\varpi^2(\Omega-\omega)^2)^{1/2}$. 

There is another way to express this. Return to Eq.~(\ref{eq:powerflux}) and introduce the quantity $d{\dot{\mathbb E}}^{\rm M}\equiv d{\dot{\mathbb E}}^{\rm B}-\Omega d{\dot {\mathbb J}}^{\rm B}$. This is also conserved along a flux surface and equals zero\footnote{In a MHD treatment there will be contributions from the conducting plasma which will render it non-zero.}. This conservation law is satisfied all the way from the horizon to the load despite the need for physical observers to move through the M frame beyond the outer light surface and within the inner light surface. The shear stress, $\{T^{\rm M\,{\hat\theta}{\hat\phi}},T^{\rm M\,{\hat r}{\hat\phi}}\}=\{B_{\rm \hat r},B_{\rm\hat\theta}\}B_{{\hat\phi}}/4\pi$, is responsible for the transport of the non-zero, conserved, rotational angular momentum flux.

If we transform into the Z basis, the electric field is
\begin{equation}\label{eq:Zelecf}
{\bf E}^{\rm Z}=\frac{(\omega-\Omega)}{2\pi\alpha}{\tilde\nabla}\Phi,
\end{equation}
while the magnetic field is
\begin{equation}\label{eq:Zmagf}
{\bf B}^{\rm Z}=-\frac1{2\pi\varpi}{\vec e}_{\hat\phi}\times{\tilde\nabla}\Phi-\frac I{\alpha\varpi}{\vec e}_{\hat\phi}
\end{equation}

\subsection{Powering the Jet}\label{ssec:energyhorizon}

To extract rotational energy, $\Omega$ must be smaller than $\omega_{\rm ILS}$ (the frame dragging potential at the inner light surface). At the outer light surface $\Omega>\omega_{\rm OLS}$. This implies that ${\bf E}^{\rm Z}$ reverses sign between the two light surfaces.
In that case, ZAMOs will measure a Poynting power per unit magnetic flux $d{\dot{\mathbb E}}^{\rm Z}=\alpha^{-2}(\Omega-\omega)I$  \citep[][hereafter BZ]{blandford77}.
This has also been shown by numerical simulations \citep[e.g.][]{koide2002, komissarov2005}. This is outward at large radius where $\Omega>\omega$ and inward at small radius where the inequality is reversed.

Waves that transport the energy outward along the field must be able to escape the ergoregion. This is a consequence of the shape of the characteristics, and hence, on the shape of the poloidal magnetic field. We discuss this in the next two sections.

\subsubsection{Causality}\label{sssec:causal}
Although we have so far described stationary, axisymmetric solutions, real magnetospheres will vary in azimuth and time. Even if these variations die off, the asymptotic solutions still reference the causal properties  of the full, time-dependent solution. These are expressed by two pairs, (forward and backward), of small local high frequency waves in the M frame. There is a ``Fast'', F, mode with electric perturbation along the direction of ${\bf k}\times{\bf B}$. This propagates locally, without currents, in the same way as a vacuum electromagnetic wave at the speed of light, although the plasma is necessary for propagation in a varying background. In addition, there is an ``Alfv\'en'', A, mode with magnetic perturbation along ${\bf k}\times{\bf B}$. This mode, which involves current, transports energy along the direction of the background magnetic field.

A proper discussion of the solution of the full time-dependent equation requires a careful specification of the boundary conditions and a mathematical definition of the characteristic surfaces. For our purposes, it is helpful to consider a 1D configuration, where we imagine two rigid conducting surfaces following neighbouring flux surfaces, separated by $\Delta\Phi<<\Phi$, and where we suppose that the electromagnetic field becomes and remains  axisymmetric.  We then need only consider variation with poloidal distance along the surfaces and time. This is a proxy for satisfying the transverse equation of force-balance (Sec.~\ref{sssec:stressbal}). 

Let's start with the light surfaces which we can think of as the critical surfaces associated with the A mode.
The relevant 1D characteristics can be thought of as short wavelength waves propagating along the poloidal magnetic field direction. The outer light surface then defines the boundary where a small, inward-directed disturbance can still propagate inward along a flux surface. As such, it is a vestige of the initial conditions. Likewise, for the inner light surface and an outward-directed disturbance. At the outer light surface, the ZAMO electric field is ${\hat{\bphi}}\times {\bf B}^Z$ with opposite sign at the inner light surface. The M-frame toroidal magnetic field is $-I/\alpha\varpi$ and varies smoothly through these surfaces. No extra condition is needed at these surfaces to allow the electromagnetic energy to flow continuously through them. By contrast, the M-frame poloidal field will tend to zero as the light surfaces are approached. 

Now, turn to the event horizon, which we can think of as the critical surface associated with the F mode. For observers falling inward across the horizon, the electromagnetic field will be finite with ${\cal I}\equiv B^2-E^2>0$.  
The acceleration relative to the  freely-falling observer diverges as the horizon is approached, as also does the relative radial 4-velocity. The ZAMO will therefore measure perpendicular, transverse field components, $E^{\rm Z}_{\hat\theta}\sim(\Omega_{\rm H}-\Omega)\Phi_{,\,\theta}/2\pi\alpha\rho$, $B^{\rm Z}_{\hat\phi}\sim-I/\varpi\alpha$ that become equal and also diverge as the horizon is approached. The radial magnetic field, $B_{\hat r\,{\rm H}}^{\rm Z}$, is unchanged by the ZAMO acceleration. This implies, 
\begin{equation}\label{eq:inflow}
I=(\Omega_H-\Omega)\varpi_{\rm H}^2B_{\hat r\,{\rm H}}^Z=\frac{(\Omega_H-\Omega)\varpi_{\rm H}\Phi_{,\theta}}{2\pi\rho_{\rm H}},
\end{equation}
evaluated at the horizon, is an inflow boundary condition \citep[cf][]{znajek77}.

Next, we consider the electromagnetic field, well beyond the outer light surface. Electromagnetic disturbances can communicate partially ``upstream'' through F modes in our 1D problem. Well beyond the outer light surface, $\varpi>>1/\Omega$, the toroidal magnetic field will dominate the poloidal component, $B_{\rm p}$ and tend to  the (poloidal) electric field. This implies that
\begin{equation}\label{eq:outflow}
I=\Omega\varpi^2B_{\rm p}=\frac{\Omega\varpi|{\tilde\nabla}\Phi|}{2\pi}.
\end{equation}

The force-free condition will start to fail after the point where this outflow condition, Eq.~(\ref{eq:outflow}), can be imposed in the load. 

There are at least three failure modes. Firstly, there can be a non-zero value for ${\bf E}\cdot{\bf j}$ which will lead to heating of and radiation by accompanying plasma. This dissipation involves an irreversible entropy production. Secondly, there can be an acceleration of the plasma by the electromagnetic force density. This is non-dissipative. Thirdly there can be entrainment of gas through the walls of the jet. The end result will be an outflow that is more reasonably described by relativistic MHD (Sec.~\ref{ssec:basequ}). 

\subsubsection{ElectroMotive Force}
A helpful way to think about these two boundary conditions is to associate an effective resistance with the horizon and with the Outer Super-Alfv\'enic Zone, OSAZ, beyond the OLS. Consider two neighbouring flux surfaces separated by flux $d\Phi$. They also have a difference in voltage $dV=\Omega d\Phi/2\pi$. $dV$ and the current $I$ are conserved through the horizon. Eq.~(\ref{eq:inflow}) can be expressed in the form $dV/I=\Omega d\Phi/2\pi(\Omega_{\rm H}-\Omega)\varpi_{\rm H}^2B_{\hat r\,{\rm H}}$. The right hand side looks like resistance --- the units are $30\,{\rm Ohm}$ --- but is not because the current, when it completes within the horizon flows from low to high $V$\footnote{It is possible to associate a resistivity, $4\pi\equiv377\,{\rm Ohm}$, with the horizon, by imagining that the current completes within it as viewed by a ZAMO. There is also a well-defined EMF $\Omega_{\rm H} d\Phi/2\pi$ associated with the two flux surfaces. Of this, a fraction $1-\Omega/\Omega_{\rm H}$ is associated with the potential difference across the horizon and a fraction $\Omega/\Omega_{\rm H}$ is associated with the load \citep[cf.][]{thorne86}.}. This relation should instead be seen as a functional relationship between $V$ and $I$ that is maintained along the flux surface to the OSAZ where $dV/I=d\Phi/2\pi r^2\sin^2\theta B_{\rm p}$ (Eq.~(\ref{eq:outflow}). Here the current will flow from high to low V in the load, although it is not necessarily dissipated in a resistance as we have just explained. 

So, by equating $dV/I$ at the horizon and the OSAZ, we can solve for $\Omega(\Phi)$ and $I(\Phi)$ if we assume shapes for the flux surfaces. Typically, $\Omega\sim\Omega_{\rm H}/2$ and $I\sim\Omega_{\rm H}\Phi/4\pi$. The energy and angular momentum flows can then  be computed using Eq.~(\ref{eq:powerflux}). Typically, comparable powers $\sim\Omega_{\rm H}^2\Phi^2/16\pi^2$ will be transported to the load and dissipated within the horizon, increasing the irreducible mass of the hole. In this 1D model, these powers are determined by conditions at the horizon and in the outflow, which, in turn, are causally determined by the electromagnetic field between the light surfaces.

\subsubsection{Stress Balance}\label{sssec:stressbal}
We next remove the conducting surfaces in our ID model, retaining stationarity and axisymmetry. This allows us to nest a sequence of ID models with the current outflow on the outside of one interval $\Delta\Phi$ almost balancing the current inflow in the next interval. This requires that there be stress balance across the surfaces and, therefore, throughout the jet. This condition determines the shape of the flux surfaces across the jet. The flux itself must be retained ultimately by the orbiting gas in the disc. For the moment we introduce a single rigid conducting surface of specified shape that crosses either the horizon or the equatorial plane. This allows us to describe how a fixed quantity of flux is distributed within this single, bounding surface. 

The most straightforward way to impose stress balance is to project the divergence of the stress-energy tensor perpendicular to a flux surface. However, a more instructive approach, leading to the same conclusion, is to start from the 4-current vector $\vec j$. Its divergence vanishes and we can relate its poloidal components in B coordinates to the scalar poloidal current $I(\Phi)$ by
\begin{equation}\label{eq:curdef}
\{j^r,j^\theta\}=-\frac1{4\pi(-g)^{1/2}}\frac{dI}{d\Phi}\{\partial_\theta\Phi,-\partial_r\Phi\},
\end{equation}
imposing stationarity and axisymmetry\footnote{A helpful way to understand this equation is to rewrite the radial component in the form $dI=-2(\rho j^r/\Delta^{1/2})\alpha(2\pi\varpi\rho d\theta)$. The 2 includes both hemispheres and recalls that $I$ represents current flowing into the hole. The first parentheses bracket the radial current density measured in a local orthonormal basis. The $\alpha$ converts charge per unit proper time to charge per unit time at infinity. The second parentheses contain the element of area.}. Now, take the poloidal part of the force-free equation, Eq.~(\ref{eq:ffc}) and use Eq.~(\ref{eq:FB}) to substitute for components of the field tensor to obtain
\begin{equation}\label{eq:RGS1}
j^\phi-\Omega j^t=\frac I{2\alpha^2\varpi^2}\frac{dI}{d\Phi}.
\end{equation}
The final step is to use Eq.~(\ref{eq:vecj}) again to substitute for the azimuthal current density, $j^\phi$ and the charge density, $j^t$, to obtain the general relativistic generalization of the Grad-Shafranov, Scharlemann-Wagoner equation for a stationary, axisymmetric solution
\citep{grad1958, shafranov1966, scharlemann73},
\begin{equation}\label{eq:RGS2}
{\vec\nabla}\cdot\left(\frac{\beta^2}{\alpha^2\varpi^2}
{\vec\nabla}\Phi\right)+\frac{(\Omega-\omega)}{\alpha^2}\frac{d\Omega}{d\Phi}{\vec\nabla}\Phi\cdot{\vec\nabla}\Phi+\frac{2\pi^2}{\alpha^2\varpi^2}\frac{d I^2}{d\Phi}=0,
\end{equation}
subject to suitable boundary conditions. The first and second terms include the combined stress from the poloidal electric and magnetic fields; the third term adds the toroidal magnetic field.

It is useful to re-express Eq.~(\ref{eq:RGS2}) as a variational principle, \citep[see][for the special relativistic case]{uchida1997b}. It is easier to consider this as an exercise in the calculus of variations than one in electrodynamics. We introduce a functional ${\cal F}(\Phi,\Phi_{,r},\Phi_{,\theta},r,\theta)$ so that 
$\delta\int drd\theta(-g)^{1/2}{\cal F}=0$. A suitable choice is
\begin{equation}\label{eq:functional}
{\cal F}=\left(\frac1{2\pi\varpi}\right)^2 {\vec\nabla}\Phi\cdot{\vec\nabla}\Phi-\left(\frac{\Omega-\omega}{2\pi\alpha}\right)^2 {\vec\nabla}\Phi\cdot{\vec\nabla}\Phi-\left(\frac I{\alpha\varpi}\right)^2.
\end{equation}
We note that we can rewrite $\cal F$ in terms of the poloidal and toroidal magnetic and electric field components in Z.
\begin{equation}\label{eq:vpBE}
{\cal F}=B^{Z\,2}_p-E^{Z\,2}_p-B^{Z\,2}_\phi={\cal I}-2B^{Z\,2}_\phi.
\end{equation}
The relationship to the Lagrangian description of electromagnetic field is of interest and will be discussed further in Paper II.

Eq.~(\ref{eq:functional}) allows us to obtain approximate solutions for flux surfaces threading the horizon. We choose a functional form $\Phi(r,\theta; p_i)$ with a few adjustable parameters, $p_i$. We then use the inflow and outflow conditions, Eqs.~(\ref{eq:inflow},\ref{eq:outflow}), to solve for $\Omega(\Phi), I(\Phi)$. Next we evaluate the functional $\cal F$ in terms of $p_i$ and minimize it by varying these parameters so as to obtain an approximate solution. Note that the inflow condition guarantees that the integral converges at the horizon and the outflow condition ensures convergence at large distance.

There is another way to look at this problem.  We have derived an approximate outflow boundary condition by supposing that the poloidal electric field is comparable with the toroidal magnetic field beyond the outer light surface. The associated resistance per unit magnetic flux should be constant along a given flux surface. This, in turn, implies a nonlinear PDE which can be solved numerically to obtain the shape of the outgoing flux surfaces. Ultimately, this is an expression of electromagnetic stress balance.

Using these solutions for $I(\Phi)$ and $\Omega(\Phi)$, we can then determine the total power and torque extracted directly from the hole by integrating Eq.~(\ref{eq:powerflux}) the over the horizon.
\begin{equation}\label{eq:powertorque}
{\dot E}=\int_0^{\pi/2}d\theta\frac{\Phi_{,\theta}^2\Omega(\Omega_{\rm H}-\Omega)\varpi}{4\pi^2\rho},\ 
{\dot J}=\int_0^{\pi/2}d\theta\frac{\Phi_{,\theta}^2(\Omega_{\rm H}-\Omega)\varpi}{4\pi^2\rho}.
\end{equation}

\subsection{Illustrative Solution}\label{ssec:bhillus}
Continuing with our illustrative solution (Sec.~\ref{ssec:illustris}), we should estimate the strength of the magnetic field between the light surfaces, above the horizon as this sets a scale for the poloidal magnetic field strength for the entire ergomagnetosphere. If all of the jet power, $L_{\rm {jet}}\sim6\times10^{43}\,{\rm erg\,s}^{-1}\sim\Omega_{\rm H}^2\Phi_{\rm H}^2/16\pi^2$, derives from the horizon, then $\Phi_{\rm H}\sim10^{33}\,{\rm G\,cm}^2$, $V\sim2\times10^{19}\,{\rm V}$, $I\sim6\times10^{17}\,{\rm A}$. The shape of the flux surfaces depends upon the boundary conditions. For the range of possibilities we envisage, the poloidal field strength will be in the range $30-100\,{\rm G}$, and the corresponding magnetic stress $\sim100-1000\,{\rm dyne\,cm}^{-2}$.

\begin{figure*}
\centering
\includegraphics[scale=0.15]{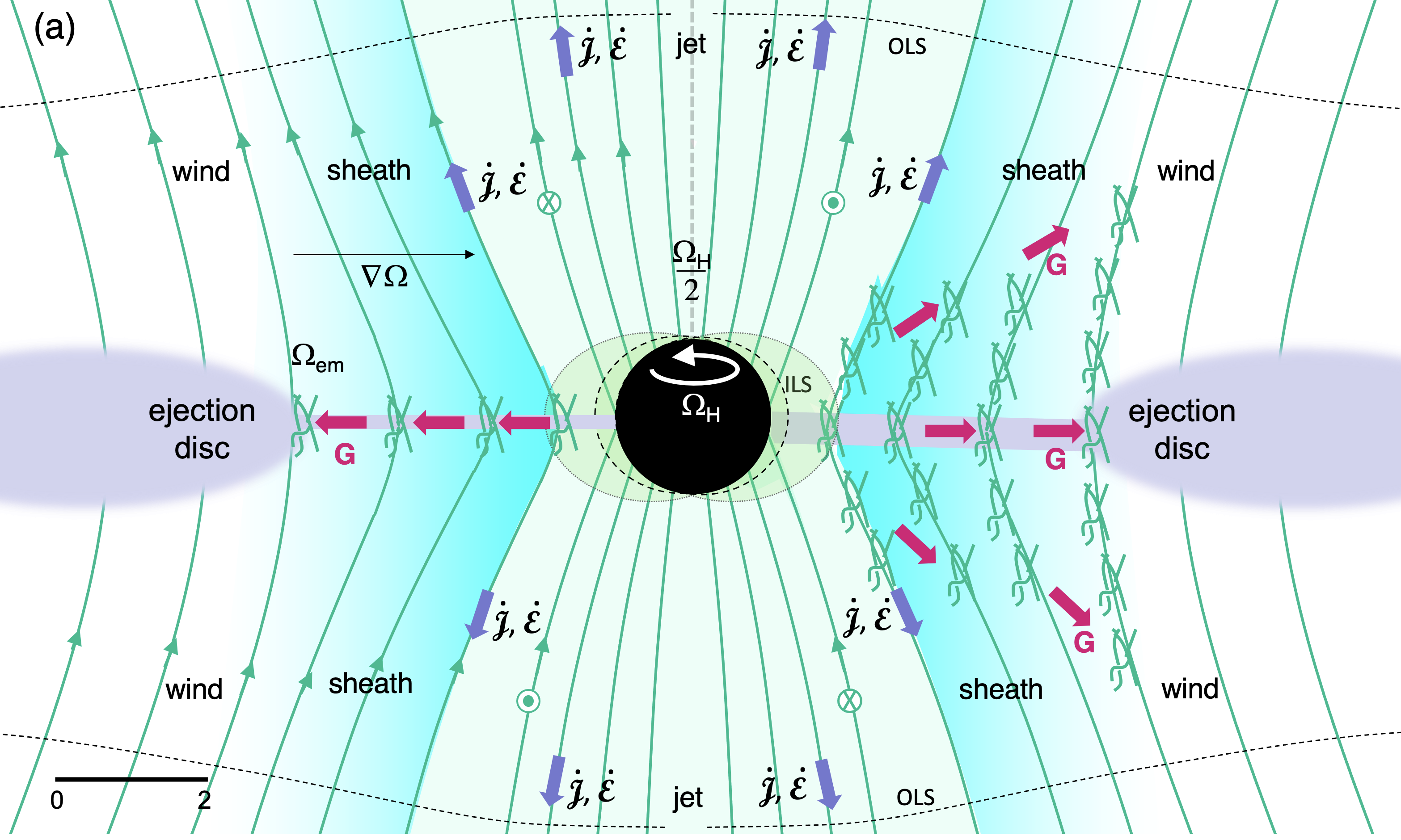}
\includegraphics[scale=0.15]{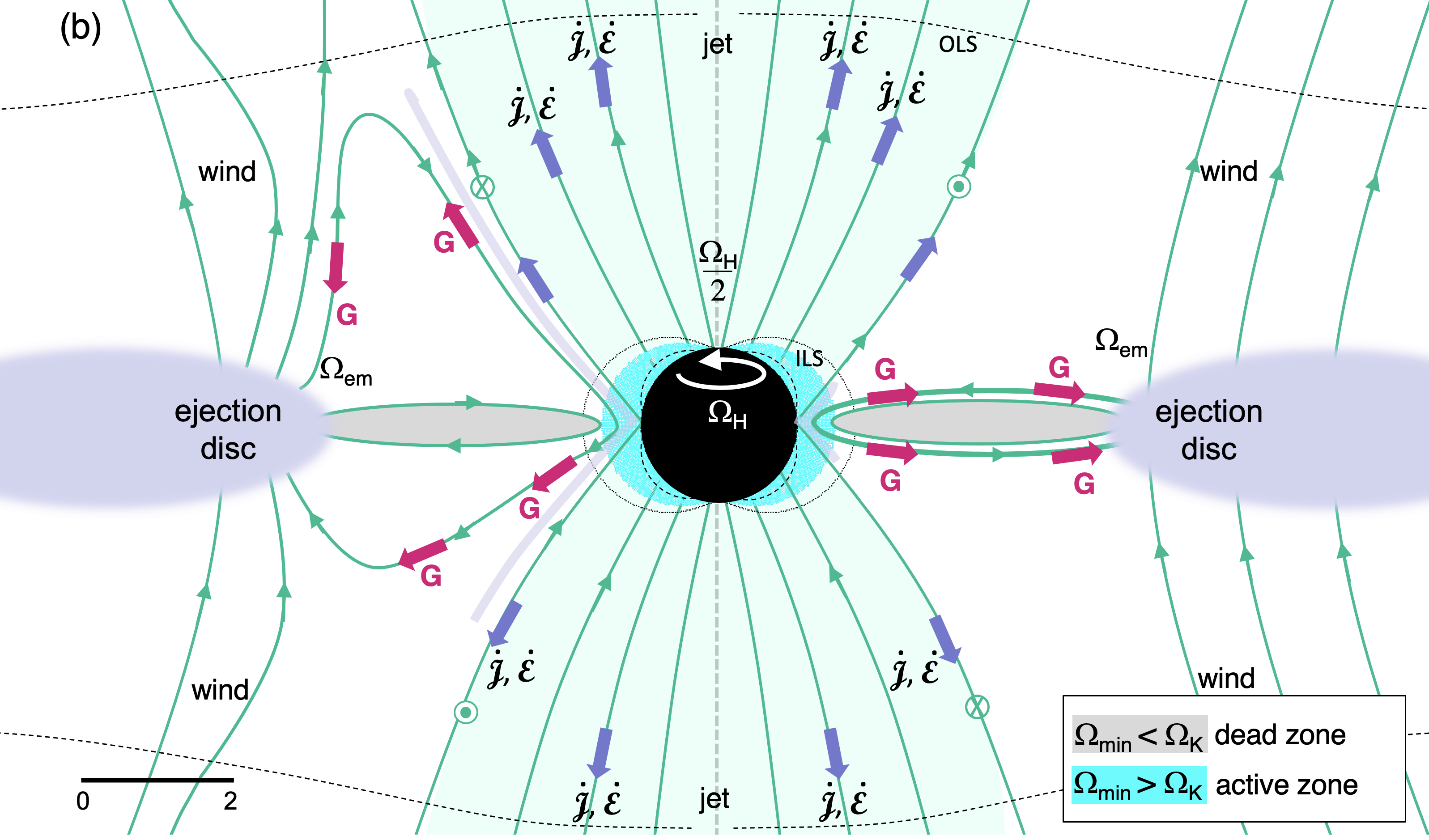}
\caption{Ergomagnetosphere within $r\lesssim 10$. Two alternative magnetic geometries are considered, presuming that mass flow is low. a) Electromagnetic clutch mechanism, where magnetic flux threads the equatorial plane of the ergosphere and, consequently, develops a non-zero $\Omega$ (Sec.~\ref{ssec:magclutch}). Non-axisymmetric, interchange instabilities transport angular momentum across the flux surfaces from the flux threading the horizon eventually to the ejection disc. On the left hand side, it is envisaged that the electromagnetic disturbances are low amplitude and that $E\rightarrow B$ in ergosphere, precipitating the governor mechanism (Sec.~\ref{ssec:governor}) where electromagnetic disturbances (F modes) are generated in negative energy orbits. On the right hand side, more extensive instabilities develop and the energy transfer to the disc will be larger. b) Magnetic capstan mechanism, where flux loops, attached to the inner disc, are continuously created threading the ergosphere and perhaps the horizon (Sec.~\ref{ssec:capstan}). Magnetic flux loops that exist outside the ergosphere will form a dead zone and just rotate with the angular velocity of the disc. Potentially unstable current sheets are likely to form. On the left hand, the flux loops are large while on the right hand side, they are confined to low latitude. In practice, all three mechanisms may operate simultaneously within a strongly turbulent, non-axisymmetric, unsteady ergomagnetosphere. Also shown are the inner, ILS, and outer, OLS, light sources which define the region from which the jets are launched.}
\label{fig:ergomag}
\end{figure*}

\section{Ergomagnetosphere}\label{sec:ergomagnetosphere}
In Sect.~\ref{ssec:energyhorizon}, we have discussed how rotational energy can be extracted by magnetic flux threading the hole horizon. In the following, we consider the region comprising field lines threading not only the hole horizon, but also  the ergosphere. We show that these field lines enable transport of energy and angular momentum beyond the outer light surface, through a mechanism we call "electromagnetic governor"\footnote{This is by analogy with mechanical governors which are used to keep the rate of rotation of a wheel or shaft at a fixed rate.}. 

However, we have conjectured that energy and angular momentum are also transported {\it laterally} from the black hole, through the {\it ergomagnetosphere},  to the inner boundary of the ejection disc near $\varpi_{\rm em}$ through a combination of magnetic torques and instabilities.

As we have discussed in Sec.~\ref{sec:disc}, the transition from electromagnetic dominance in the ergomagnetosphere to matter dominance in the ejection disc occurs at $\varpi_{\rm em}$. It is also the place where three inner boundary conditions can be imposed on the ejection disc.

The first boundary condition involves the mass flow. We have assumed that $\dot M\propto \varpi^n$.  This implies that there is still an appreciable flow of plasma at $\varpi_{\rm em}$. It is much larger than the supply of particles needed to carry charge while, by definition, dynamically unimportant. 
It is not clear what happens to this mass flow. One limiting case is that all of it proceeds inward to the horizon; the other is that it is essentially all ejected from this point along a surface that eventually becomes the jet sheath.  For simplicity, we adopt the ejection limit here so that we can continue to adopt the force-free electrodynamics formalism (although a MHD description with negligible inertia does not change the principles discussed below).  

The second boundary condition involves the angular momentum. In the inflow limit, we need a torque $G_{\rm em}$ applied to disc by the ergomagnetosphere to partly oppose the angular momentum carried inward by mass flow in the disc. In the ejection limit, a larger torque will be required to drive away the inflowing gas from this radius as well. However, most of the angular momentum in the ejection disc derives from large radius and $G_{\rm em}$ is irrelevant to the global angular momentum balance. The third, energy, condition is similar. It provides the source of power that ultimately drives away the mass flow at large radius and dominates the energy budget. It will be given by $\sim G_{\rm em}\Omega_{\rm em}$, the rate at which the torque $G_{\rm em}$ does work.  

We now discuss some ways by which this transfer of energy and angular momentum through the ergomagnetosphere can happen. We have identified  basic mechanisms to energize the disc which we call the ``clutch'' and the ``capstan''\footnote{With traditional automotive and nautical counterparts.}.  While we fully expect that the electromagnetic field in the ergomagnetosphere is time-dependent and non-axisymmetric and may contain appreciable plasma, as we discuss further in Paper II, it is still helpful to consider axisymmetric force-free idealizations to demonstrate general mechanisms that could be at work and also to suggest equilibria which can be used for exploring stability. 

\subsection{Electromagnetic Governor and Jet Energization}\label{ssec:governor}
Let us start with stationary, axisymmetric magnetic flux surfaces that pass through the ergomagnetosphere and the equatorial plane. This is a different geometry from that found by e.g. \citet[][]{komissarov2004, parfrey19} where an equatorial current sheet forms allowing this magnetic flux to pass through the horizon. A possible reason for this difference is that there is a problem with having vertical magnetic field pass through the ergosphere. If the flux surfaces have finite angular velocity, $\Omega$, then they will carry energy and angular momentum upwards within a sheath that surrounds the black hole jet (Fig.~\ref{fig:ergomag}). However, under our assumptions and unlike with the magnetic flux that threads the horizon, there is no ready source for this energy and angular momentum and so the flow of angular momentum, given by $I$, and the angular momentum density, given by $\Omega$, should both rapidly decrease. When $\Omega$ falls below  $\Omega_{\rm min}$, $E^Z$ can become so large that ${\cal I}\rightarrow0$. Using Eqs.~(\ref{eq:Zelecf},\ref{eq:Zmagf}), the actual transition occurs when ${\cal I}=0$ or, equivalently,\footnote{As with Eq.~(\ref{eq:inflow}), we can either think about this condition as a functional relation between $V$ and $I$ that is maintained along flux surfaces or as a version of Ohm's law requiring a frame transformation, as with a Faraday disc.} 
\begin{equation}\label{eq:etrans}
I=\frac{\{(\Omega_{\rm max}-\Omega)(\Omega_{\rm min}-\Omega)\}^{1/2}\varpi|{\tilde\nabla}\Phi|}{2\pi}.
\end{equation}
Adopting our sign convention, the current should be positive.

In this limit, normal, force-free electrodynamics becomes invalid. As discussed by several authors \citep[e.g.,][]{uchida1997, komissarov2002,komissarov2004,paschalidis2013, pfeiffer2013}, the reason for this change is that, when ${\cal I}<0$, the force free equations can no longer be written in a symmetric, hyperbolic form \citep{pfeiffer2013} and cannot be evolved in a predictable manner, for example in a simulation.
Equivalently, a Lorentz frame moving with speed $v_{\rm elec}$ in which the electromagnetic field is purely electrostatic appears. What has been widely presumed is that $\cal I$ has a floor of zero and when it is reached there is associated and sufficient pair production and $\gamma-$ray emission to allow energy to be conserved. 
However, the appearance of an {\it electric zone} with $E>B$, is a dynamic process and the transformation speed $v_{\rm elec}$ decelerates from the speed of light with large spatial and temporal gradients.
As we argue in Blandford, {\textit in prep}. and Paper II, the local partial differential equations of force-free electrodynamics should remain valid and result in a rapid growth of force-free F and A modes with wavelengths on all scales.

Under these circumstances, most of the energy no longer associated with the macroscopic, slowly varying electromagnetic field is present in these modes. Although the analogy is not perfect, this is similar to dynamical chaos or a transition to turbulence in a fluid flow when a critical Reynolds' number is reached. Under these circumstances, solutions of the force free equations with neighboring initial conditions, will diverge rapidly and are essentially unpredictable. Alternatively, we can think of this as a sort of phase transition from an ordered to a disordered state. The key difference between what is asserted here and what is presumed in many simulations is that, although there must be an inner scale where either plasma effects or QED leads to dissipation, the power dissipated locally in this fashion can be negligibly small in much the same way that when a flow becomes turbulent, the bulk kinetic energy of the original, laminar flow is carried mostly by the kinetic energy of the turbulence, not heat. 

In general, the partition into F and A modes depends upon how the condition ${\cal I}=0$ is approached. In the present case, the orientation of the large scale electric and magnetic field, measured in a limiting Z frame, ensure that F modes are generated propagating in the $-{\vec e}_{\hat\phi}$ direction. Although these wave modes are locally similar to vacuum electromagnetic modes on null geodesics, they are force-free and require current to keep them in their state of polarization when the background changes slowly under WKB conditions. The trajectories on which they are launched correspond to maximally negative energy orbits and must cross the horizon. By carrying negative energy in the ergosphere near the equatorial plane, these waves allow the magnetic flux surfaces leaving the ergosphere to carry off energy and angular momentum continuously. This is, of course, a manifestation of the Penrose process \citep{penrose1969}. It is similar to the mechanism identified by \cite{parfrey19} with the important difference that the modes can have wavelengths $\lesssim10^{14}\,{\rm cm}$ approaching the gravitational radius instead of those of the highest energy $\gamma-$rays with wavelengths $\sim10^{-14}\,{\rm cm}$. However, despite the electromagnetically-dominant conditions envisaged here, these waves should still follow null geodesics and behave like photons. The viability and relevance of this mechanism needs further discussion. 

We, therefore propose that when magnetic flux threads the ergosphere, it is not dragged into the horizon, but, instead, adopts an angular frequency that is intermediate between 0 and $\Omega_{\rm min}$, in much the same way that field lines crossing the horizon achieve an angular frequency intermediate between $\Omega_{\rm H}$ and 0. In this way, energy and angular momentum can be steadily extracted from the black hole along magnetic field lines that thread the ergosphere but not the horizon \citep[e.g.,][]{Takahashi1990, koide2003}. When the large scale electromagnetic field leads to an increased removal of angular momentum, the rate of wave production will increase to keep $\Omega\sim\Omega_{\rm min}$ and {\sl vice versa}. This mechanism provides the governor that keeps that magnetic flux rotating with an angular velocity close to the minimum allowed in the equatorial plane.

The actual boundary condition that must be applied at the governor near the equatorial plane is essentially ${\cal I}=0$. Evaluating in the Z frame, we equate the electric field from Eq.~(\ref{eq:Zelecf}) to the absolute magnetic field from Eq.~(\ref{eq:Zmagf}). When we combine this relation with the outflow boundary condition Eq.~(\ref{eq:outflow}) we can solve for $\Omega$ and $I$, just as we did for the magnetic flux threading the horizon. This allows us to solve for the energy and angular momentum flow along magnetic flux surfaces that thread the ergosphere.

What about the field lines that cross the equator outside the ergosphere? If we apply the considerations that led to the electromagnetic governor, then we would expect them to become non-rotating. Equivalently the radial current which is allowed to flow, through the ergosphere from the horizon along the equator can no longer flow outward and, so, must form a return current that will ultimately complete along the surface of the jets, forming a "sheath" (see Fig.~\ref{fig:ergomag}). In its idealized form, the electromagnetic governor only energizes the jets and not the disc. To achieve this we invoke a second, idealized mechanism, the electromagnetic clutch. 

\subsection{Electromagnetic Clutch and Disc Energization}\label{ssec:magclutch}
\subsubsection{Differential Rotation and Interchange  Instability}\label{sssec:interch}
The electromagnetic clutch operates in analogy to a mechanical clutch that transmits torque between two surfaces that rotate with different angular frequencies. It is invoked to carry energy and angular momentum laterally across open flux surfaces (left panel of Fig.~\ref{fig:ergomag}). We suppose that this clutch connects the black hole horizon and, perhaps the electromagnetic governor, to the disc. We have seen how the angular velocity of these flux surfaces and the current they carry is determined by the boundary conditions. In general we expect the angular velocity of these field lines to decrease monotonically with cylindrical radius $\varpi$ from the horizon, where $\Omega\sim\Omega_{\rm H}$ to the disc at $\varpi_{\rm em}$, where the angular velocity is that of the disc $\Omega_{\rm em}$, the associated Keplerian angular velocity.

This differential rotation of the magnetic surfaces is possible because the magnetic field is not line-tied to orbiting gas, as also pictured in Fig.~\ref{fig:ergomag}. This configuration is likely to be unstable to local interchange instabilities in which neighboring flux tubes become entangled.  This  creates a local mean shear stress communicating angular momentum across, not just along, the magnetic field \citep[e.g.,][]{yuan19}. This component of torque $dG$ will do work at a rate $\Omega dG$ and thereby transport energy outward as well. This conjectured instability is similar to the magnetorotational instability that operates when weakly magnetized, conducting plasma follows Keplerian orbits within an accretion disc \citep{balbus98,arayagochez02} with the important difference that it involves force-free electrodynamics rather than MHD, as we discuss in Paper II. Linear modes will grow on a dynamical timescale. The turbulence growth might be limited and this will be investigated in Paper II. In the Fig.~\ref{fig:ergomag} we show the two limits. On the left-hand side, we show the limit where the mean field is stronger than the fluctuations  and on the right hand side, the limit where the fluctuations and the mean field are of the same order.  

Although it is most reasonable that a clutch be located close to the equator, this is not necessary. The source of the angular momentum need only  be within the outer light surface of the hole and the ergomagnetosphere. The target associated with the disc can be anywhere in the wind below the Alfv\'en surface. Torque applied here can be communicated effectively upstream as Alfv\'enic disturbances to the disc, where most of the angular momentum lies as well as downstream to power the wind. If the clutch mechanism is effective, then it is reasonable to expect the rate of transport of angular momentum from the hole to the disc is comparable with that carried away by the jets. when this torque, $G_{\rm em}$ is applied at the disc, it will do work at a rate $G_{\rm em}\Omega_{\rm em}$. 

\subsubsection{Phenomenological Model}\label{sssec:clutchpheno}
Despite the obvious uncertainty, it is helpful to model the flow of angular momentum across a time- and azimuth-averaged ergomagnetosphere and out into the disc, phenomenologically. This is reminiscent of the challenge posed by gas dynamical accretion discs before the advent of MHD simulations, where a parameter $\alpha$ that relates the shear stress to the pressure is introduced \citep{shakura73}. In this spirit, we define a dimensionless parameter, $\aleph$, which relates the torque $G$ to the equatorial angular velocity gradient $\nabla \Omega$. Specifically we write 
\begin {equation}
G=-\aleph B_p^2\varpi^2 {\rm H}_{\rm ergo}(\varpi^{5/2} \nabla \Omega)\,,
\label{torque}
\end{equation}
where $B_p(\varpi)$ is the poloidal magnetic field which will be roughly constant and ${\rm H}_{\rm ergo}$ is the characteristic half-thickness of the ergomagnetosphere, where the torque is applied. The quantity in parentheses equals $-3/2$ for a Keplerian disc. Just as we did for the ejection disc, we can write down a conservation equation for angular momentum with the important difference that there is, by assumption, little flow of angular momentum associated with accreting gas. 

There are two sources of angular momentum for the clutch. The magnetic flux threading the horizon will interface with that threading the disc and is a likely site of instability. In addition, the governor mechanism can operate within the ergosphere and the boundary condition ${\cal I}=0$ allows to solve for the current, $I$, there. Associated with this current will be a loss of angular momentum away from the disc at a rate of ${\dot{\mathbb J}}=I/2\pi$ per unit magnetic flux. If $\aleph$ is small then there will be little, outward transport of angular momentum to the disc and the equilibrium value of $\Omega$ beyond the disc will be small.

Conversely, if $\aleph$ is large, then the outward flow of angular momentum beyond the ergosphere can be competitive with that contributed to the jet, cf. Eq.~(\ref{eq:jcon}) with $\dot M=0$. The power transmitted to the disc will then be a fraction $\sim5\aleph\Omega_{\rm em}/\Omega_{\rm H}$ of the source power, with 5 being a strongly model-dependent numerical factor. The key point is that the further out is the disc, the less power will be transferred to it.

An important consequence of having an electromagnetic clutch is that there must be dissipation associated with differential rotation, cf. Eq.~(\ref{eq:diss}). As the torque is associated with local electromagnetic interchange instabilities, it is reasonable to suppose that much of the energy liberated is carried off in F and A modes. We defer further discussion of the clutch mechanism to Paper II.

\subsection{Magnetic Capstan}\label{ssec:capstan}
\subsubsection{Magnetic Loops}

We next consider an alternative configuration of field lines in the ergomagnetosphere, which we call the magnetic capstan (right panel of Fig.~\ref{fig:ergomag}). In this case, we suppose that stationary and axisymmetric magnetic loops tied to the disc near the transition radius pass through the ergomagnetosphere and the equatorial plane, at a radius we call $\varpi_{\rm min}$, all the way down to the ergoregion. This magnetic geometry is like an inverted, toroidal magnetic arcade similar to structure occasionally seen in the solar corona that can apply a torque on the disc \citep[e.g.,][]{vanPutten1999,li2002,uzdensky2005, Krolik_2005, wang2007}. It requires the presence of two finite toroidal current sheets extending outward from the  horizon's equator. Current sheets can also form between the closed intermittent  loops and the jet walls like pictured in the left side of Figure~\ref{fig:ergomag}, panel b. Such current sheets are likely to exhibit instabilities \citep[cf.][]{mckinney2009}.

We expect that all of these magnetic loops will rotate with angular velocity $\Omega\sim\Omega_{\rm em}$, as the disc is highly conducting. Flux loops with large enough $\varpi_{\rm min}$ will satisfy the condition $\Omega_{\rm em}>\Omega_{\rm min}$ at the equator and, consequently, ${\cal I}>0$ with $I=0$, c.f. Eq.~(\ref{eq:Zmagf}). These current-free magnetic loops, that exert no torque on the disc, define what we call a ``dead zone''.  However, when $\Omega_{\rm em}>\Omega_{\rm min}$ a current, given by Eq.~(\ref{eq:etrans}),  must flow. This is what we call the "active zone" (Fig.~\ref{fig:ergomag}).  

The situation is somewhat similar to that encountered in the governor. Force-free wave modes will be created on negative energy orbits to supply the power and torque transmitted to the disc through the presence of a current $I$ determined by imposing the condition ${\cal I}=0$. The magnetic field lines will be dragged forward by the faster ergomagnetosphere ensuring that a positive torque is applied to the disc.

\subsubsection{Phenomenological Model}\label{sssec:cappheno}
We can also make a phenomenological model for the capstan. Using an estimate of the flux threading the ergosphere we obtain an estimate $\gtrsim\Omega_{\rm em}/\Omega_{\rm H}$ for the ratio of the disc power to the source power. The angular momentum carried away by the capstan is $\sim \Phi I/(2\pi)$ where
\begin{equation}\label{eq:inflow}
I=(\Omega_{\rm min}-\Omega_{\rm em})\varpi_{\rm min}^2B_{p\,{\rm min}}\,.
\end{equation}
Even adopting the assumptions of this simple model, realistic configurations have a complex structure \citep[e.g][]{yuan19}. Magnetic loops of alternate polarity could give rise to striped jets \citep[][]{Mahlmann2020, chashkina2021}. Plasma can accumulate within the dead zone and its inertia may cause loops of magnetic flux to detach and cross the horizon. There may be enough matter flowing inward that the equatorial electrodynamics is MHD, not force-free. Such developments are likely to reduce the torque applied on the disc, because the field lines become loaded with plasma which affect the efficiency \citep{globus2014}. Conversely, there may be significant flux threading the disc at one end and the horizon at the other end to increase the torque. 

An additional consideration, especially adopting the inflow boundary conditions is that we expect these loops of field to detach from the disc at regular intervals and be swallowed by the event horizon. Implicit in the capstan mechanism is a regular production of new loops of magnetic flux.

\subsection{Strongly Turbulent Regime}\label{ssec:sturb}
These three simple idealizations  demonstrate that the ergomagnetosphere can be a competitive source of power to the hole itself and that this power can both end up in the jet (including the jet sheath) and the disc. We suppose that the electromagnetic field will change greatly as the boundary conditions change. There may be transitions to  configurations with secondary field structures, analogous to the transitions seen in Couette flows \citep[e.g.][]{gorman1982}. Ultimately, the electromagnetic field may transition to a non-axisymmetric, time-dependent fully turbulent state. It may exhibit intermittent examples of the governor, clutch and capstan just as the solar corona exhibits examples of their non-relativistic, slowly rotating counterparts. The possibility of this outcome and the associated flow of angular momentum and energy must be explored by high resolution, numerical simulations such as those pioneered by \citet{parfrey19, ripperda2022} and references cited therein.

\subsection{Illustrative Solution}\label{ssec:emillus}
We continue discussion of our illustrative solution by noting that the current flowing downward along the jets, $\sim6\times10^{17}\,{\rm A}$, is much larger than the minimal current needed at $\varpi_{\rm em}$ which is $\sim10^{16}\,{\rm A}$.  This suggests that there is a large return current flowing outward and defining the jet sheath. If most of the jet current is concentrated in its core, then this allows the stress exerted by the MHD wind to be smaller than that within the center of the jet.

\section{MHD Wind and Jet collimation}\label{sec:mhdwind}
In Sec.~\ref{sec:disc}, we outlined a model of the accretion disc orbiting the black hole all the way from $\varpi_{\rm em}$, through the ejection disc, to $\varpi_{\rm ed}$, through the infall, emphasising the conservation laws of mass, angular momentum, energy and current. In this section, we pay attention to the MHD wind and discuss how it interacts with the inflowing gas from the surrounding galaxy and cluster.

We also considered the magnetization of the disc, anticipating the presence of a strong magnetic field within the disc, responsible for the internal torque, $G$, and a large scale, unipolar field responsible for driving the MHD wind which is ultimately responsible for collimating the jet.

\subsection{Launching the Wind: the Role of the Ergomagnetosphere}\label{ssec:launchw}
We first consider how the wind may be launched. At the surface of the disc, there will be closed loops of magnetic field being constantly twisted, stretched by differential rotation and squeezed upward by magnetic stress. In addition there will be open magnetic field lines passing through the disc, organised into an array of flux tubes of various sizes, with spatially separated footpoints. The differential rotation in the disc will twist them with the opposite sense to the overall angular velocity. The current within a flux tube will be parallel to the current associated with the horizon and the ergomagnetosphere but antiparallel to the local average current density. It must be over-cancelled by the individual return currents associated with the region between the flux tubes. We further suppose that the open flux tubes will merge above a few disc scale heights and form a large scale roughly stationary and axisymmetric magnetic flux distribution that it is interrupted by frequent coronal mass ejections, fast streams and interactions with stellar winds and clouds. They are also likely to mix. 

We anticipate that any gas that finds itself on a flux tube that develops a significant radial component will be flung out by centrifugal force. For many open flux tubes, this will require that gas is raised by thermal or relativistic particle pressure high enough above the disc for magnetic stress to take over. Some flux tubes will be better fed with gas than others but the feeding will change constantly. Again, we suppose that above a few disc scale heights, the gas flow will smooth out. Now in a simple model of a MHD wind, we suppose that gas on a single flux tube passes through a slow wave critical point followed by an Alfv\'en wave critical point and then a fast wave critical point. As our primary concern in this section is with what happens to the wind far from the disc, we shall ignore the flow before the slow point, and ignore the thermal pressure all together. We consider the overall wind flow over many decades of radius without presuming self-similarity.

As with our description of the force-free electrodynamics around the black hole, we shall make the important simplification that the large scale flow is stationary and axisymmetric and, again, we imagine two neighbouring flux surfaces that are held in place by imaginary rigid  conductors along these surfaces. The flow of angular momentum and energy and, now, mass are conserved between these boundaries. Given these assumptions, the specific angular momentum evaluated in the inertial frame, is given by $\ell=\varpi(v_\phi- B_\phi/4\pi{\mathbb M})$, where ${\dot{\mathbb M}}=\rho v_p/B_p=d{\dot M}/d\Phi$ is the conserved mass flux per unit magnetic flux and the subscript $p$ denotes the poloidal component. Likewise, the conserved specific energy in the inertial frame --- the Bernoulli function --- is $e=(v_p^2+v_\phi^2)/2-1/r-\Omega\varpi B_\phi/4\pi{\dot{\mathbb M}}$, where $\Omega$ is the conserved angular velocity and equal to the equatorial, Keplerian angular velocity in a thin disc at the magnetic footpoint. We next transform into a frame rotating with angular velocity $\Omega$. The magnetic field and poloidal speed are unchanged while the azimuthal speed is ${\tilde v}_\phi=v_\phi-\Omega\varpi=v_pB_\phi/B_p$ where $B$ is the total magnetic field. The speed in this frame is ${\tilde v}=(v_p^2+{\tilde v}_\phi^2)^{1/2}=v_pB/B_p$. The conserved, effective specific energy in the rotating frame is ${\tilde e}=e-\Omega\ell={\tilde v}^2/2-1/r-\Omega^2\varpi^2/2$ is the effective gravitational potential in the rotating frame, including the centrifugal potential.

In the rotating frame, the energy $\tilde e$ can be expanded as the sum of a kinetic energy ${\tilde v}^2/2$ plus an effective potential energy, ${\tilde V}_{\rm eff}$ which can be Taylor-expanded about the point at the center of the disc where $\Omega$ equals the Keplerian angular velocity and the cylindrical radius is $\varpi$. If the horizontal and vertical displacements from this point are $\delta\varpi$, $\delta z$, respectively, then 
\begin{equation}\label{eq:veff}
{\tilde V}_{\rm eff}=-\frac{3\Omega^{2/3}}2\left(1+\delta\varpi^2-\frac13\delta z^2+\dots\right).
\end{equation}
Using this effective potential, we immediately recover the result that the gas will be flung out centrifugally when the gradient of the local field line satisfies $d\delta\varpi/d\delta z>\delta\varpi/3\delta z$ or, for a straight field line passing through the Keplerian point, the angle with respect to the vertical exceeds $30^\circ$\citep{blandford82}\footnote{It can also be flung inward when $d\delta\varpi/d\delta z<\delta\varpi/3\delta z$.}. If this inequality is well-exceeded close to the disc, then the gas supply will be quickly depleted allowing the gradient to decrease. For this reason we will choose to model the launching of the wind with the boundary condition $B_\varpi/B_z=3^{-1/2}$ applied at the Keplerian footpoint. The equipotential passing through the Keplerian footpoint asymptotes to $\varpi=3^{1/2}\Omega^{-2/3}$. This suggests that the magnetic flux surfaces do not become closer to the axis than this.

\subsection{Collimating the Wind: the Role of the Magnetopause}\label{ssec:windout}
An important features of this model is that the MHD wind that is powered by the black hole spin alongside the jet is responsible for the collimation of the jet. In order for this to occur it is necessary for the wind from the ejection disc to be collimated itself. This, in turn, requires that the outer boundary of the wind enforce this collimation. As has been argued elsewhere, \citep{meyer19, blandford18}, a feature of the infall in M87 and other giant elliptical galaxies is that it is likely to be quasi-spherical. It will shock, but the momentum flux it carries is likely to be as large at high latitude as at low latitude. If, conversely, the gas supply is mostly in the equatorial plane, as in a spiral galaxy, such as our own, then the boundary conditions imposed on any outflow from a disc are likely to be quite different. The flow lines will not be forced to converge and any outflow will not be collimated.

We can implement this boundary condition by defining a converging rigid surface emanating from $\varpi_{\rm ed}$ with shape chosen to give a normal stress roughly $\propto r^{-2}$. Simultaneously, we remove all the rigid surfaces separating adjacent flux surfaces. This requires us to impose transverse force balance throughout the outflow. 

A most important feature of the outflow is the Alfv\'en surface where the total flow speed equals the local Alfv\'en speed. This is located at $\varpi=\varpi_{\rm A}=(\ell/\Omega)^{1/2}$. Unlike other critical points, an Alfv\'en point is an attractor and imposes no conditions upon an outflow that transitions from sub- to super-Alfv\'enic. However, it does designate a boundary below which all characteristic information can be transmitted downward toward the disc. We have already explained how the connection between the ergomagnetosphere and the disc can occur below the Alfv\'en surface which might be at quite high latitude. This may limit the potentially  destructive effect of rotating kink instabilities around the Bondi radius.

Similar considerations apply to the interface between the outflowing gas emanating from the ejection disc and the inflowing gas headed towards the injection disc. We call this interface the ``magnetopause'', by analogy with the terrestrial counterpart. It may be unstable and poorly defined. Changes below the Alfv\'en surface can be communicated to the disc; above the disc, only changes in the distribution of magnetic flux can be transmitted and this will only be possible below a second critical point associated with the fast wave, where the poloidal flow speed equals the Alfv\'en speed. 

An additional feature of having a high latitude magnetopause is that it is more likely to retain sufficient magnetic flux to facilitate the formation of a collimated MHD wind. in general, the flux associated with the outer disc is much larger than that near the black hole, but it still sets a scale  and so may also be conducive to the formation of powerful jets and the formation of double radio sources. A third consequence of having the magnetopause extend to high latitude is that it will likely inhibit the growth of destructive and unobserved kink instability in the inner wind flow and consequently the jet \citep{Eichler1993,spruit1997}. 

To demonstrate this, one has to consider the Grad-Shafranov equation \citep[][]{Heinemann1978,okamoto1975, lovelace1986} and impose the external pressure profile at the magnetopause that will act to collimate the disc outflow. Shocks may also form at this outer boundary. In general an MHD wind can be efficiently collimated by the external pressure of the surrounding medium. It is of interest that the differential equation for the transverse stress balance in the wind can also be expressed as a variational principle, quite analogous to Eq.~(\ref{eq:functional}). This can also be used to explore the stability of the wind and the jet it confines (Paper II).

\subsection{Collimating the Jet: the Role of the Ejection Disc}
We have proposed that the dynamical pressure of the inflow at large cylindrical radius may be transmitted through the MHD outflow from the ejection disc to the relativistic jet created by the black hole magnetosphere and, perhaps, the ergomagnetosphere. (Note that this interpretation of jet collimation is quite different from that proposed in e.g. \cite{cruzosorio22} where a thick accretion disc is presumed to extend to altitudes $\gtrsim1000$ and provided the collimation.) We now turn to the nature of this second interaction. It is clearly observed in the case of M87 beyond a distance $z\gtrsim100$ and is even better defined in other sources. As a boundary layer, it can evolve in two complementary ways. Firstly, it can be seen as an interface that begins as quite thin relative to the body of the jet across which momentum is transported by essentially viscous stress in with a fluid jet. The flow in the interface may be laminar or, more likely, turbulent but in either case it will broaden until it encompasses the entire jet. In this case, mass as well as momentum will be entrained by the jet which will decelerate it and change its equation of state.

Alternatively, the interface could be primarily electromagnetic, in which case the dissipation is primarily due to electrical resistivity and the spreading of the flow of momentum is mediated by a larger electromagnetic field. It is this interaction which can ultimately determine if a source become FR-I or FR-II, depending upon whether it is the jet or the wind that carries the greater momentum. This interpretation of jet confinement within the Bondi radius stands in marked contrast to alternative interpretations of the confining medium as the extension of a thick torus or as infalling gas.

In the case of M87, the jet cylindrical radius $\varpi_{\rm jet}(z)$ has been carefully measured, provided that we accept the value of the jet inclination. This allows us to explore how the confinement takes place. Turning first to the jet itself, there are two important limits to consider. Firstly, a jet that is dominated by relativistic plasma, with sound speed $\sim3^{-1/2}$, will not be in causal contact if its bulk Lorentz factor exceeds the reciprocal of the jet opening angle times a model-dependent number $O(1)$. Alternatively, if, as we are implicitly assuming here, the body of the jet, within the boundary layer, is primarily electromagnetic then stress waves can cross it in an expansion timescale and it can be considered as being in internal stress balance. 

If we assume that the electromagnetic field inside this boundary layer naturally evolves from a well-organized form close to the black hole, albeit allowing for some imperfect internal dynamics, then the toroidal magnetic field component in the fame of the source will vary as $B_\phi\propto I(\varpi_{\rm jet})\varpi_{\rm jet}^{-1}$. The poloidal magnetic field will vary roughly according to $B_z\sim\Phi_{\rm jet}\varpi_{\rm jet}^{-2}$. However, there will also be poloidal electric field whose stress will partly compensate the magnetic stress. The electric field will vanish and the toroidal magnetic field will diminish  within a relativistic outflow frame while the poloidal magnetic field is unchanged. Such an electromagnetic configuration may be resilient to destructive instability, as observed.

Turning to the jet-wind interface, we anticipate the presence of weak, oblique shocks in the wind \citep[][hereafter GL16]{globus16}. The component of the total stress normal to the interface should balance across a contact discontinuity. This shear layer is likely to be unstable and to lead to mixing of slower moving gas into the jet --- the ``viscous'' part --- and faster moving electromagnetic field --- the ``resistive'' part --- into the wind, a recipe for efficient particle acceleration by a variety of mechanisms as observed. The entire  mixing region --- the sheath --- will also broaden until it meets in the center of the jet, perhaps in a Mach disc or a recollimation shock, which it is tempting to associate with HST-1 \citep{levinson2016}.

Given the presence of these shocks, the shape of the interface will be determined by the relevant normal components of momentum flux. For the MHD wind this includes the bulk motion and a magnetic component, mostly associated with $B_\phi$. If we suppose that a centrifugally driven wind cannot get closer to the axis than its footpoint, (Sec.~\ref{ssec:launchw}), then $\varpi_{\rm jet}(z)$ must increase monotonically with $z$.

\citet{lyubarsky2009} analyzed the force-free GS equations and showed that beyond the Alfv\'en surface, a parabolic force-free jet requires the presence of ambient medium with a pressure profile  $p_{\rm ext} \propto z^{-2}$ in the collimation zone.  Such a profile can be produced by an extended disc wind, and GL16 showed that efficient collimation occurs when  the wind power $\gtrsim0.1$ of the jet power. Therefore, if we follow this simple prescription, the torque imposed by the ergomagnetosphere on the disc must perform work at a rate $G\Omega_{\rm em}= L_{\rm wind} \sim 0.1 L_{\rm jet}$ to collimate the jet. However GL16 only consider the pressure of the wind shocked layer and ignored the effect of the toroidal field. Here, we have a different boundary condition. In the injection disc, the field is mostly poloidal. In the ejection disc, the wind will build up toroidal field. Now, this wind carrying away mass and confining the force-free spine jet,  is embedded in the large scale poloidal field built up by the infall (i.e., the poloidal field dominates over the toroidal field for $\varpi>\varpi_{\rm ed}$) and also, by the pressure of the external medium at the Bondi radius.

\section{Summary and Discussion}\label{sec:discuss}
In this paper, we have a developed an alternative interpretation of recent observations of M87 that have imaged the region around its 6.5 billion solar mass black hole and the innermost parts of the limb-brightened jets it produces. We have argued that, just as the observations span scales from the event horizon $\sim10^{15}\,{\rm cm}$ to those larger than the Bondi radius, $\gtrsim10^{21}\,{\rm cm}$, our understanding must involve a comparable range of scales connected by conservation laws associated with mass, angular momentum, energy and electrical current. We propose that the major power source for both the jets and the disc of infalling gas, is not gravitational energy but rotational energy of the black hole. In other words, in sources like M87, ``nature'' is more important than ``nurture''.

More specifically, this hypothesis has the consequence that almost all of the gas that joins the disc around the Bondi radius is ejected from smaller  radii $\sim10^{16-20}\,{\rm cm}$ with a natural feedback allowing just enough mass to accumulate and magnetic flux to be trapped near the black hole to allow outward transport of energy through a relatively cool and thin disc to power the outflow. (This is analogous to the sun finding a stable equilibrium where its central temperature is high enough to allow nuclear reactions to compensate its loss of energy by radiation.) This model, which is an extension of the ideas presented in \cite{blandford99,1999ASPC..160..265B}, is very different from the MAD/MCAF models and other simulations of accretion discs where there is a conservative flow of mass through a hot, yet radiatively inefficient, torus. The other difference is that our model invokes an ergomagnetosphere to transfer efficiently a fraction of the rotational power from the spinning black hole to the disc.

We have argued that the outflow is likely to take the form of a MHD wind in which cold gas is flung out centrifugally. The wind is carrying away most of the mass over a range of radii in contrast to the results of simulations \citep{Zhu2018}. This wind is supposed to be confined laterally by infalling gas at high latitude. Again, we expect that there is a global feedback mechanism which lets the mass loss adjust locally so that sufficient power flows outward through the disc to allow it to escape. With the wind, a combination of toroidal magnetic field and bulk flow transmit this stress to small cylindrical radius providing active collimation of the jets.

Four interfaces are a major feature of this interpretation (shown in Fig.~\ref{fig:Fig1}). Firstly, there is the magnetopause which separates the infall from the disc wind on a scale which we suppose is $\varpi_{\rm ed}\sim10^{20}\,{\rm cm}$. The interface  may be observable.

Secondly, there is the hot corona associated with the cooler disc, where gas is launched onto the wind. The corona must be hot enough to launch the gas, averaging over large enough area and time because the detailed magnetic structure of the corona must evolve on an orbital timescale. The torque associated with the disc and the corona will also adjust on this timescale and this will feedback into the dissipation responsible for the heating of the corona and the disc emission. 

Thirdly, there is the ergomagnetosphere where electromagnetic power from the spinning black hole  \citep[estimated as $\sim0.1$ times the total jet power, as needed to collimate the jet as estimated by][]{globus16}, is transported to the inner disc at $\varpi_{\rm em}\sim5\times10^{15}\,{\rm cm}$. Here we have identified several basic mechanisms, involving interchange instabilities or direct magnetic connection, may feature in a strongly varying and possibly turbulent region. We have outlined how the presence of electric zones, where $E\rightarrow B$, may lead to the production of waves on negative energy orbits that can allow power to be extracted from the hole by magnetic flux threading the magnetosphere but not the event horizon. 

Fourthly, there is the observed interface between the jet and the wind --- the sheath --- observed at de-projected distances $\gtrsim 5\times10^{17}\,{\rm cm}$ \citep[][and references therein]{Hada2019}. This is an electromagnetic boundary layer where generalized viscous and resistive dissipation takes place and entrainment, particle acceleration and radio emission follow\footnote{Note that a limb-brightened jet structure has recently been observed in Cen A \citep{janssen2021}.}. Understanding each of these interfaces is essential for determining the overall flow of mass, energy, angular momentum, energy and current as well as the feedback mechanisms at work that allow M87 to operate on timescales, supposedly much longer than the infall rate at the Bondi radius. It is also central to making comparison with possible future observations. These include making simple models of the mm observations intensity and polarisation assuming the emission derives from the ergomagnetosphere and the sheath. In addition, the wind contains a dense cold plasma and an organized magnetic field which can be a source of Faraday rotation and depolarisation, especially for lines of sight that pass through the sheath. Finally the hypothesized magnetopause and the outer disc might be observable by JWST. We defer a fuller discussion of observational implications to Paper III.

Although we have focused on M87 in this paper, it is of interest to ask how the same physical considerations might pertain to other types of source. We suspect that this type of flow is generic in slowly accreting, massive black holes in elliptical galaxies and leads to double radio sources unless the black hole spin is low. The relevance to radio quasars where the mass accretion rate is presumably much higher \citep{krolik1999} might involve a large mass flow rate of radiation-dominated gas and similar dynamical considerations might be relevant, albeit under quite different physical conditions.
In jets associated with X-ray binaries, it has been shown that powerful, pressure-driven jet can emanate from the accretion disc \citep[e.g.][]{Ferreira2006I}.
 
Similar remarks apply to microquasars where electromagnetic power is likely to dominate fluid power \citep[see e.g.][]{Middleton2014}, or jets associated with gamma-ray bursts (GRBs). Neutrino heating has been invoked as a possible mechanism to power the jets, if the mass accretion rate is sufficient to turn on nuclear reactions in the inner edge of the accretion disc. Subsequent creation of a hot outflow of made  electron-positron pairs, due to neutrino anti-neutrino annihilation above the horizon, can produce jets with luminosities as high as $\sim 10^{52}$ erg~s$^{-1}$ \citep{Zalamea2011}. However the highest  luminosity reachable through this process is still not sufficient to explain the most energetic GRBs observed, which are likely to be powered by the spinning black hole if the magnetic flux near the horizon is high enough \citep[e.g.][]{globus2014b}. 

It is hoped that the remarkable observations of M87 by the EHT collaboration, which stimulated this investigation, will be continued at shorter wavelength, longer baselines and with different sources, notably Sgr A$^\ast$, and that this study provides a useful, alternative framework for their interpretation.

\section*{Acknowledgements}
RB thanks St. Johns College, Oxford University, the Solvay Institute, Caltech and KITP for hospitality.
NG's research is supported by the Simons Foundation, The Chancellor Fellowship at UCSC and the Vera Rubin Presidential Chair. NG and RB acknowledge the National Science Foundation under Grant No. NSF PHY-1748958 to KITP. The authors thank past collaborators, especially those cited below, for their ideas, encouragement and critique over many years. 

\section*{Data Availability}
No new data were generated or analysed in support of this research.

\appendix
\bibliographystyle{mnras}
\bibliography{ref} 
\section{Zero Torque Solution}\label{sec:zerot}
It is instructive to consider a mathematically possible, though physically unrealistic, solution where the torque within the disc is small enough to be ignored, $G=0$\footnote{In the spirit of the Carnot cycle.}. From Eq.~(\ref{eq:jmdot}), there is no dissipation\footnote{There is ohmic dissipation associated with the inward flow of gas across a stationary magnetic field. However, this is negligible for a thin disc.} and no radiation. The disc is invisible. The absence of heating implies that the gas must be flung away cold and invisibly from the ejection disc.  This requires a MHD wind which applies an external torque to the disc, causing mass to flow inward at a rate $\dot M$ and to leave the disc at a rate, per $\ln\varpi$, $\dot{\cal M}$.

Let us consider the flow of mass, angular momentum and energy in this flow adopting the formalism of Sec.~\ref{ssec:claw} and setting the internal torque, $G$, to zero. The inward flow of angular momentum in the disc, $-\dot J$, is given by $\dot M\ell_{\rm K}$. The divergence of this flow is the angular momentum carried off by the wind which is the sum of a gas part $\dot{\cal J}_{\rm g}=\dot{\cal M}\ell_{\rm K}$ and a magnetic part $\dot{\cal J}_{\rm m}=\frac12\dot M\ell_{\rm K}$. The outward flow of energy is ${\dot E}=\frac12{\dot M}\Omega_{\rm K}\ell_{\rm K}$. (The mass flow is inward and the energy is negative.). Its divergence is the energy carried off by the wind which is also the sum of a gas part $\dot{\cal E}_{\rm g}=-\frac12\dot{\cal M}\Omega_{\rm K}\ell_{\rm K}$ and a magnetic part $\dot{\cal E}_{\rm m}=\Omega_{\rm K}\dot{\cal J}_{\rm m}=\frac12\dot M\Omega_{\rm K}\ell_{\rm K}$.

The total energy carried off by the wind $\dot{\cal E}=\dot{\cal E}_{\rm g}+\dot{\cal E}_{\rm m}$ should be positive if the wind is to escape to infinity.  This implies that the mass cannot be lost too rapidly from the disc, specifically that $0\le d\ln M/d\ln\varpi<1$.

\section{Transformations}\label{sec:cotrans}

The Boyer-Lindquist coordinate system suffices to describe the electromagnetic field, despite the fact that a physical observer on a timelike geodesic within the ergosphere cannot remain with constant $r,\theta,\phi$. However it is instructive to transform, into a local Lorentz "M" frame, the Magnetic frame, orbiting the hole at fixed $r,\theta$ with fixed electromagnetic angular velocity $\Omega$. We will only be interested in the coordinates of physical quantities measured in a local, orthonormal basis with basis vectors $\vec{\bf e}^{\rm M}_{\hat\alpha}\equiv\{\vec{\bf e}_{\hat t},\vec{\bf e}_{\hat r},\vec{\bf e}_{\hat\theta},\vec{\bf e}_{\hat\phi}\}$. The transformation matrix for transforming a contravariant vector components forwards from B to M, is given by 
\begin{equation}\label{eq:transBO}
L^{{\rm BM}\hat\alpha}_{\qquad\beta}=
\begin{pmatrix}
\frac1\beta&0&0&\frac{(\Omega-\omega)\varpi}{\alpha\beta}\\
0&\frac{\Delta^{1/2}}\rho&0&0\\
0&0&\frac1\rho&0\\
\frac\Omega\beta&0&0&\frac{\beta^2+\Omega(\Omega-\omega)\varpi^2}{\alpha\beta\varpi}
\end{pmatrix}
\end{equation} 
where $\beta=(\alpha^2-(\Omega-\omega)^2\varpi^2)^{1/2}$. The inverse matrix transforms the basis vectors forwards, covariant components forwards and contravariant components backwards. 

We see from Eq.~(\ref{eq:transBO}) that the transformation becomes singular when $\beta\rightarrow0$. This has two, complementary, interpretations. Firstly a fixed $r,\theta$ observer, orbiting with the electromagnetic angular velocity $\Omega$, will only be on a timelike geodesic so long as $\Omega_{\rm min}=\omega-\frac\alpha\varpi<\Omega<\omega+\frac\alpha\varpi=\Omega_{\rm max}$. Secondly, a set of such observers sharing the same field line will only be on timelike geodesics if they are located between an inner light surface where $\Omega_{\rm min}=\Omega$ and an outer light surface where $\Omega_{\rm max}=\Omega$. Physical observers within the inner light surface must move inward towards the horizon; those beyond the outer light surface must move outward towards the load.

By contrast, the motion of a ZAMO will always be timelike outside the horizon because $\Omega_{\rm min}<\omega<\Omega_{\rm max}$ there. The corresponding transformation matrix from the B frame to an orthonormal ZAMO or Z basis, with basis vectors $\vec{\bf e}^{\rm Z}_{\hat\alpha}\equiv\{\vec{\bf e}_{\hat t},\vec{\bf e}_{\hat r},\vec{\bf e}_{\hat\theta},\vec{\bf e}_{\hat\phi}\}$, is
\begin{equation}\label{eq:transBZ}
L^{{\rm BZ}\hat\alpha}_{\qquad\beta}=
\begin{pmatrix}
\frac1\alpha&0&0&0\\
0&\frac{\Delta^{1/2}}\rho&0&0\\
0&0&\frac1\rho&0\\
\frac\omega\alpha&0&0&\frac1\varpi
\end{pmatrix}
\end{equation} 
\end{document}